\documentclass[sigconf]{acmart}
\AtBeginDocument{%
  \providecommand\BibTeX{{%
    \normalfont B\kern-0.5em{\scshape i\kern-0.25em b}\kern-0.8em\TeX}}}

\setcopyright{rightsretained} 
\copyrightyear{2024} 
\acmYear{2024} 
\acmDOI{10.1145/3613904.3642446}
\acmConference[CHI '24]{Proceedings of the CHI Conference on Human Factors in Computing Systems}{May 11--16, 2024}{Honolulu, HI, USA}
\acmBooktitle{Proceedings of the CHI Conference on Human Factors in Computing Systems (CHI '24), May 11--16, 2024, Honolulu, HI, USA}
\acmISBN{979-8-4007-0330-0/24/05}

\usepackage{multirow}
\usepackage{verbatim}
\usepackage[table]{colortbl}
\usepackage{xcolor}
\usepackage{booktabs,caption}
\usepackage{setspace}
\usepackage{listings}
\usepackage{acmart-taps}
\usepackage{cancel}
\usepackage{tabularx}
\usepackage[ruled,vlined,linesnumbered,lined]{algorithm2e}

\SetNlSty{textbf}{}{:} 
\newcounter{cpModel}
\makeatletter
\newenvironment{cpModel}[1][!htbp]{%
  \stepcounter{cpModel} 
  \begin{algorithm}[#1]
  \DontPrintSemicolon 
  \SetAlgoLined 
}{
  \end{algorithm}
}
\makeatother

\begin{document}

\title{Evaluating the Utility of Conformal Prediction Sets for AI-Advised Image Labeling}

\author{Dongping Zhang}
\email{dzhang@u.northwestern.edu}
\orcid{0000-0001-9825-1411}
\affiliation{%
  \institution{Northwestern University}
  \city{Evanston}
  \state{Illinois}
  \country{USA}
}

\author{Angelos Chatzimparmpas}
\email{angelos.chatzimparmpas@northwestern.edu}
\orcid{0000-0002-9079-2376}
\affiliation{%
  \institution{Northwestern University}
  \city{Evanston}
  \state{Illinois}
  \country{USA}
}

\author{Negar Kamali}
\email{negar.kamali@u.northwestern.edu}
\orcid{0000-0002-1086-6735}
\affiliation{%
  \institution{Northwestern University}
  \city{Evanston}
  \state{Illinois}
  \country{USA}
}

\author{Jessica Hullman}
\email{jhullman@northwestern.edu}
\orcid{0000-0001-6826-3550}
\affiliation{%
  \institution{Northwestern University}
  \city{Evanston}
  \state{Illinois}
  \country{USA}
}

\renewcommand{\shortauthors}{Zhang et al.}




\begin{abstract}
As deep neural networks are more commonly deployed in high-stakes domains, their black-box nature makes uncertainty quantification challenging. We investigate the presentation of conformal prediction sets---a distribution-free class of methods for generating prediction sets with specified coverage---to express uncertainty in AI-advised decision-making. Through a large online experiment, we compare the utility of conformal prediction sets to displays of Top-$1$ and Top-$k$ predictions for AI-advised image labeling. In a pre-registered analysis, we find that the utility of prediction sets for accuracy varies with the difficulty of the task: while they result in accuracy on par with or less than Top-$1$ and Top-$k$ displays for easy images, prediction sets offer some advantage in assisting humans in labeling out-of-distribution (OOD) images in the setting that we studied, especially when the set size is small. Our results empirically pinpoint practical challenges of conformal prediction sets and provide implications on how to incorporate them for real-world decision-making.
\end{abstract}





\maketitle

\section{Introduction}\label{sec:intro}

As machine learning models such as deep neural networks (NNs) achieve impressive predictive performance, AI-advised decision-making has become more commonplace in various situations where a human must make a decision, from high-stakes domains like medicine (e.g.,~\cite{razzak2018deep}) or autonomous vehicle navigation (e.g.,~\cite{fujiyoshi2019deep}) to online tasks like content moderation (e.g.,~\cite{lai2022human}) or image labeling (e.g.,~\cite{wal2021biological}). Ideally, having access to a predictive model can help humans make more informed decisions, surpassing the capabilities of either human judgment or AI in isolation, a synergy known as \textit{complementary performance}~\cite{bansal2021does, hemmer2021human}. Whenever an AI system understands certain regions of the feature space better, the human counterpart can benefit by considering the model predictions. For example, in AI-advised image labeling, which we study in this paper, human labelers may benefit from viewing predictions from a model trained on previously labeled examples.

To help humans know when to trust model predictions, presenting information about the probability that the model's prediction is correct can, in principle, be useful. For example, a decision-maker could be provided with the softmax pseudo-probability produced in the final layer of NNs for classification tasks. However, NN predictions are known to be overconfident at times~\cite{guo2017calibration}, and heuristic measures of uncertainty are often poorly calibrated. Consequently, decision-making relying on NN predictions with the highest softmax scores (e.g., Top-$1$ or Top-$k$) can fail to account for prediction error, leading to worse decisions. 

Given these challenges, conformal prediction~\cite{vovk2005algorithmic, romano2020, angelopoulos2020uncertainty} has emerged as an alternative solution that can rigorously quantify the prediction uncertainty for NNs through the use of \textit{prediction sets}. Instead of presenting unreliable softmax values or other heuristics for uncertainty, the prediction set is a type of confidence interval consisting of a set of predictions with a \textit{coverage guarantee}---the true class is, on average, captured within such sets with a user-specified probability (e.g., $95\%$). The conformal framework is compatible with NNs because the derived sets are \textit{distribution-free}---they are model-agnostic, work with finite samples, and offer non-asymptotic guarantees without distributional assumptions~\cite{angelopoulos2021gentle}. 

One challenge is that while conformal prediction sets express prediction uncertainty with guarantees that status quo presentations like Top-$k$ predictions with softmax cannot, the conformal \textit{coverage guarantee} can lead to large sets embodying high uncertainty for instances the model perceives as difficult~\cite{babbar2022,angelopoulos2020uncertainty}. Compared with Top-$k$ predictions, such sets may increase cognitive load, rendering uncertainty quantification less effective~\cite{zhou2017effects}, thus compromising predictive performance. To understand how conformal prediction sets support human decisions aided by model predictions in an image labeling task, we contribute a large online repeated measure experiment ($n=600$) that compares the accuracy achieved in AI-advised image labeling using conformal prediction sets with that achieved with no predictions or Top-$k$ predictions. Participants in our study were tasked with labeling images from ILSVRC 2012~\cite{deng2009imagenet} that varied in difficulty and encompassed both image stimuli that were in-distribution and out-of-distribution (OOD). This allowed us to evaluate the utility of prediction sets against Top-$k$ presentation alternatives in scenarios where the model's predictions were mostly accurate versus when the presence of ``unknown unknowns'' makes predictions more error-prone. 

We evaluate decision-making using metrics such as \textit{accuracy} and the \textit{shortest path length}, which approximates the amount of deviation from the ground truth in the label space hierarchy. We use participants' elicited \textit{willingness-to-pay} at the end of the experiment to compare the perceived value of predictions against the post-hoc inferred monetary benefits derived from prediction access.

We find that for in-distribution instances, prediction sets lead to reduced labeling accuracy compared to Top-$k$ predictions. However, for OOD instances where model misspecification differentially affected the calibration of Top-$k$ predictions, prediction sets improve accuracy regardless of the set size. We find that participants are willing to pay roughly equivalent amounts for each type of display; if anything, they undervalue prediction sets relative to other presentations. Our results advance understanding of the strengths and limitations of prediction sets for improving AI-advised decision-making and highlight decision-makers' tendency to over-rely on predictions, even when poorly calibrated, in an image labeling task.
\section{Background}\label{sec:relwork}
\subsection{Uncertainty Quantification Using Conformal Prediction}
Prior work on Human-Computer Interaction (HCI) suggests that humans can make more informed decisions when uncertainty is effectively communicated (e.g., ~\cite{joslyn2012uncertainty,kay2016ish,hullman2018pursuit}), such as by presenting prediction uncertainty through static intervals~\cite{cumming2014new,cumming2005inference,manski2019communicating,taylor1994guidelines} or animations~\cite{hullman2015hypothetical, zhang2021visualizing}. However, uncertainty quantification for NNs remains challenging~\cite{amodei2016concrete,angelopoulos2021gentle} due to their black-box nature and unreliability of internal heuristic confidence values, such as softmax pseudo-probabilities~\cite{guo2017calibration, torralba2011unbiased}. 

One approach to quantify prediction uncertainty in classification tasks is using Bayesian NNs, where predictions depend on sampling from the posterior distributions of model parameters, carrying an inherent, mathematically grounded measure of uncertainty. However, Bayesian NNs heavily rely on distributional assumptions for their priors and can be computationally intensive to train. \textit{Conformal prediction}~\cite{vovk2005algorithmic} has emerged as a more flexible and efficient solution that does not require strong model or distributional assumptions. In classification settings, conformal prediction quantifies uncertainty through \textit{prediction sets}. Similar to traditional confidence intervals, the derived sets contain predicted classes and achieve a \textit{coverage guarantee} that, on average, the true class is captured by the set with a user-specified probability. Formally, given a training set $(x_1, y_1), (x_2, y_2), \dots, (x_n, y_n)$, where $x_i$ is the feature vector and $y_i$ is the corresponding label, the uncertainty set function, denoted $C(X)$, maps a feature vector $x_i$ to a subset of $Y = \{1, \dots, K\}$. This mapping ensures \textit{coverage}\footnote{Recent work explores alternatives to marginal coverage, including class-conditional coverage~\cite{gibbs2023conformal} and relaxed versions of conditional coverage (e.g.,~\cite{hore2023conformal}).} that, for any new instance $x$ drawing from the same distribution as the training data, the probability $P(Y \in C(X)) \geq 1 - \alpha$ over the randomness in the calibration and test points, with an error rate of $\alpha$~\cite{angelopoulos2021gentle}.

If an uncertainty quantification approach is well-calibrated, we naturally expect it to reflect greater uncertainty on more difficult instances for the model (e.g., OOD instances). With prediction sets~\cite{romano2020, angelopoulos2020uncertainty}, this manifests as \textit{adaptiveness}: a conformal prediction set will be larger when the task is more difficult for the model. Adaptiveness is realized through the inductive conformal method, which quantifies the prediction uncertainty of a chosen classifier using a designated \textit{calibration set}~\cite{papadopoulos2002inductive} separate from the training data set. This set contains instances that are independently and identically distributed (i.i.d.) to those on which the model is trained.

For each calibration instance, the predicted softmax scores are adjusted using Platt scaling~\cite{platt1999probabilistic, guo2017calibration, nixon2019measuring} for better interpretability and further regularized to penalize unlikely classes~\cite{angelopoulos2020uncertainty}. These calibrated probabilities are then ranked in descending order, reflecting the classifier's confidence in each class. A conformity score is computed by summing these ranked probabilities up to and including the true class, capturing the model's accumulated confidence for the instance leading up to the correct classification. From the distribution of the conformity scores of all calibration instances, a threshold is computed by taking the $\alpha$ quantile after a finite sample correction for robustness. For a new and unknown instance, this calibration threshold decides the size of the prediction set by including all labels whose softmax scores exceed this calibration threshold. In our study, we opted for the Regularized Adaptive Prediction Sets (RAPS) algorithm~\cite{angelopoulos2020uncertainty} to ensure that the prediction sets of our experiment are as small as possible while achieving, on average, the desired coverage rate.

\subsection{Effects of Presenting Uncertainty in AI-advised Decisions}
Pairing humans with an AI system in complex decision-making scenarios can elevate performance beyond what either can achieve individually when their strengths are complementary~\cite{bansal2021does,guo2024statistical, steyvers2022bayesian}. In annotation tasks with large label spaces, AI models can improve accuracy and efficiency~\cite{levy2021assessing}. Particularly, in AI-advised image labeling, various tools exist for applying automation to enhance human annotators' speed and accuracy~\cite{sager2021A}. However, performance on AI-advised tasks can be influenced by how predictions are presented---specifically, whether uncertainty is effectively communicated or not. Effective communication of uncertainty can improve the perceived trustworthiness of predictions~\cite{zhang2020effect, lai2019human, beller2013improving}. Presenting a model's accuracy is necessary to provide humans with a well-defined decision problem in many AI-advised decision settings~\cite{hullman2024decision}, and can encourage analytical thinking and reduce over-reliance on predictions~\cite{zhang2020effect, prabhudesai2023understanding}. However, uncertainty quantification does not guarantee improved decision-making. Cognitive biases, such as tendencies to under- or over-react to information relative to rational standards, play a major role in how uncertainty is interpreted (e.g.,~\cite{ambuehl2018belief,tversky1971belief}). Additionally, uncertainty quantification can improve trust under low cognitive load, but can diminish trust when uncertainty is presented ambiguously, demanding high cognitive load~\cite{zhou2017effects}. 

There is limited empirical evidence on how prediction sets influence trust and accuracy in AI-advised decision-making. One exception is \citet{babbar2022}, who investigated humans' perceived utility and performance using RAPS versus Top-$1$ predictions for labeling images from CIFAR-100~\cite{krizhevsky2009learning}. They found that participants using RAPS reported higher trust and perceived utility than those using Top-$1$ predictions. However, their study was unlikely to have achieved sufficient power to detect differences in accuracy between the two groups due to the very small sample size (15 participants per group). They suggest that smaller sets might lead to better accuracy, but this observation was based on a comparison between sets generated by two different conformal methods. Recent work by \citet{straitouri2023} compared labeling performance between two decision support systems: one that only allowed participants to select a label value from the prediction set and another that allowed them to freely choose one of 16 labels. Their results suggest that when the model is well-calibrated, decision-making solely based on prediction sets tends to be more reliable and accurate because of the coverage guarantee.

Motivated by overlapping research questions to \citet{babbar2022}, we conducted a much larger experiment to evaluate the utility of different uncertainty prediction displays for in-distribution and OOD stimuli, each systematically categorized by the difficulty and size of the associated prediction set. This enabled us to explore changes in performance by instance type. Contrary to the restricted agency in~\citet{straitouri2023}, our study allowed participants to freely navigate a large label space, so as to allow for the possibility that even inaccurate predictions may serve as useful clues that help participants deduce the correct answer.
\section{Online Experiment}\label{sec:exper}
\subsection{Overview}
\begin{figure*}[ht]
  \centering
  \includegraphics[width=\linewidth]{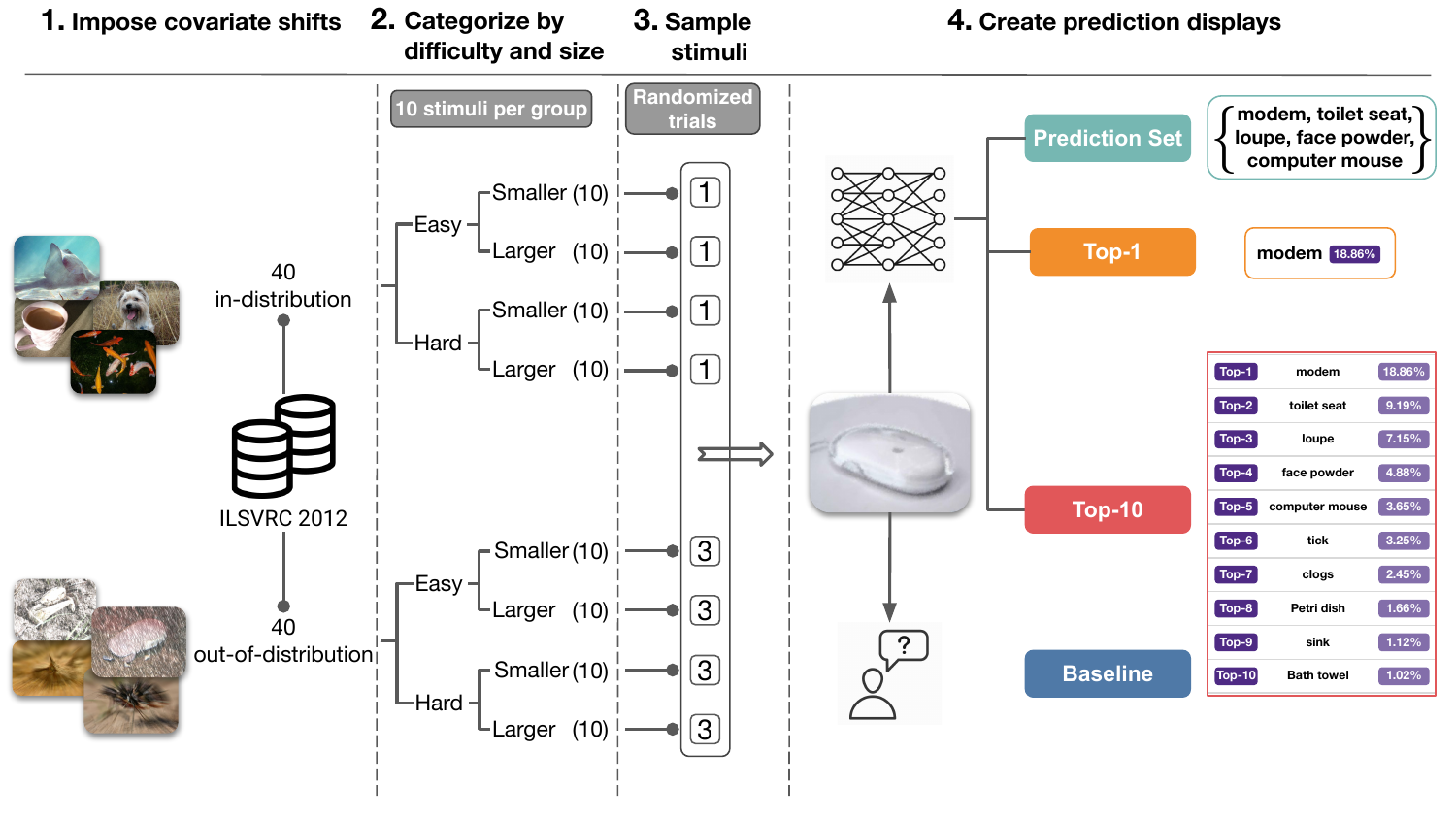}
  \caption[]{
  Overview diagram of our key experimental manipulations. (1) Five different covariate shifts are imposed through synthetic image corruption to create five replications of the \textit{conformal hold-out set}, each containing images that are OOD. (2) Images in each conformal hold-out set are categorized by the classifier's prediction confidence for difficulty and the size of the derived prediction set. Ten task images representative of the categories used to define each group are selected. (3) Participants label 16 task images sampled from 80 candidate images: four in-distribution and 12 OOD, balanced by difficulty and set size, presented in randomized order. Example task stimuli are shown in \autoref{fig:stimuli-example}. (4) Based on the conditions assigned, participants may complete labeling tasks without predictions (i.e., \textcolor[HTML]{4D79A7}{\textbf{\texttt{baseline}}}) or with access to prediction displays that vary in the content provided by uncertainty quantification (i.e., \textcolor[HTML]{F28E2B}{\textbf{\texttt{Top-1}}}, \textcolor[HTML]{E15759}{\textbf{\texttt{Top-10}}}, or \textcolor[HTML]{77B7B3}{\textbf{\texttt{RAPS}}}). Screenshots of the interface as seen by participants are presented in \autoref{fig:interface-display}.
  }
  \label{fig:exp-design}
  \Description{
  This figure is an overview diagram illustrating the key experimental manipulations used in the study. It visualizes the process of utilizing conformal hold-out sets to create OOD images through synthetic image corruption, categorization of these images by difficulty and set size, and the subsequent selection of task images for the labeling tasks. The diagram further depicts the interface for different treatment conditions, including labeling without predictions (e.g., the baseline condition) or with access to various prediction displays with varying uncertainty quantification, across a series of tasks involving both in-distribution and OOD images.
  }
\end{figure*}

We conducted a large mixed-design repeated measure experiment on Prolific. Following a pre-registered\footnote{Pre-registration: \url{https://aspredicted.org/6gh27.pdf}} analysis plan, we evaluated how different displays of prediction uncertainty impacted performance on an AI-advised labeling task, including relative to performance without access to predictions. Figure~\ref{fig:exp-design} provides an overview of our stimuli generation process, which systematically varied whether image instances were in-distribution or OOD and easy or difficult, as well as the size of conformal prediction sets.

Each participant in our study labeled a set of 16 images sampled from the ILSVRC 2012. We assigned participants to one of four \textit{conditions} representing different \textit{prediction displays}: (1) no predictions (baseline), (2) Top-$1$ prediction with softmax, (3) Top-$10$ predictions with softmax, or (4) the RAPS prediction set~\cite{angelopoulos2020uncertainty}. 
 
In real-world scenarios, the deployed model can encounter unexpected OOD instances where the distribution of input features differs from what was trained and calibrated. We varied whether images were in-distribution or OOD within subjects. Among the 16 stimuli, 4 were in distribution, while 12 were not, as we expected greater variation in participants' accuracy for OOD. 

In our experiment, OOD instances notably lowered the model's Top-$k$ prediction accuracy. Although the corrupted images based on the ILSVRC 2012 dataset that we used as OOD instances could compromise the  conformal coverage guarantees, the adaptiveness of RAPS led to good coverage despite OOD. However, the number of labels included in the prediction sets became much larger when the instance was harder for the model. To evaluate the impact of set size on the effectiveness of prediction sets, we balanced set size within subject by categorizing images as smaller or larger based on the size of the prediction set generated for the image. This factor enabled us to examine the performance of the prediction set between size levels, as previous work suggests that a smaller set size is more preferable~\cite{romano2020, angelopoulos2020uncertainty} and may yield higher accuracy for in-distribution images~\cite{babbar2022}. We expected smaller set sizes to be more useful in general, as they could reduce cognitive load while maintaining the conformal coverage guarantee for in-distribution images. Larger sets embodying high uncertainty could still be useful for labeling when participants were highly uncertain of what the image showed and consequently heavily relied on prediction sets. In our setting, set size could also be viewed as a nuanced measure of difficulty for in-distribution stimuli because images that required a larger set size to achieve the desired coverage were intrinsically more difficult~\cite{angelopoulos2020uncertainty}.

For both in-distribution and OOD stimuli, we balanced difficulty within subject by easy and hard instances, using the median cross-entropy loss of in-distribution images as a threshold to define these levels. The former enabled us to assess whether participants were inclined to accept predictions that expressed high confidence. The latter allowed us to examine whether participants were more likely to modify or reject low-confidence predictions and instead rely on their own intuition.

Our study evaluated participants' performance by their labeling \textit{accuracy} and the \textit{shortest path length} between the chosen label and the correct label in the WordNet hierarchy~\cite{miller1995wordnet, fellbaum1998wordnet}. After they had labeled all images, we asked them to report their \textit{willingness-to-pay} for the prediction display they used (i.e., a value ranging from $0$ to $8$ USD, the maximum they could earn) if they were to repeat the study with a similar distribution of instances. Finally, we elicited the strategies employed to approach the tasks.

\subsection{Task and Stimuli Generation}\label{sec:dgm}

\subsubsection{Overview of Stimuli Generation}
To generate stimuli, we sourced 80 task images from ILSVRC 2012, evenly divided between in-distribution and OOD (i.e., 40 of each). For both in-distribution and OOD images, we further categorized each set of stimuli by difficulty (between \texttt{easy} and \texttt{hard}) and by set size (between \texttt{smaller} and \texttt{larger}). Note that because the range of observed set sizes differed depending on the difficulty of the instance, set sizes were defined relatively rather than absolutely. This categorization resulted in four distinct groups, each comprising ten images, for both \texttt{in-distribution} and \texttt{OOD} categories. Image stimuli within each group were selected to exemplify their respective categories, determined by the model's (described below) prediction confidence and the size of the prediction set derived.

\subsubsection{Predictive Model for Image Classification}
Given that underperforming models are rarely deployed in real-world scenarios, we chose a classifier designed to achieve high prediction accuracy. We used a pre-trained Wide Residual Network (WRN)~\cite{zagoruyko2016wide} on ILSVRC 2012. The WRN is a Residual Network (ResNet) variant~\cite{he2016deep}, and its primary architectural distinction is the widening factor in the channel width of the convolutional layers compared to a ResNet of the same depth. Specifically, we utilized a \texttt{wide\_resnet101\_2} model with $101$ layers and a widening factor of $2$, using PyTorch's \texttt{IMAGENET1K\_V2} weights~\cite{paszke2019pytorch}. This model was well-calibrated for in-distribution instances, achieving a Top-$1$ accuracy of $82.5\%$ and a Top-$10$ accuracy of $97.9\%$ over all in-distribution images, as detailed in \autoref{tab:all-corruptions-top1-ranking} of Appendix~\ref{sec:appendix-methodol}. We used the WRN to create Top-$1$ and Top-$10$ predictions along with their corresponding softmax pseudo-probabilities to establish prediction displays for Top-$k$ conditions.

\subsubsection{Calibration Set}
To produce prediction sets, we randomly sampled half of the 50,000 images from the ILSVRC 2012 validation set to create a \textit{conformal calibration set}, which contained 25,000 in-distribution images that were unseen by the model. This calibration set was used to perform \textit{inductive conformalization} by calibrating the model based on the conformity scores derived from all images in the \textit{conformal calibration set}. This process enabled the creation of prediction sets for the remaining 25,000 images in the \textit{conformal hold-out set}. 
Our prediction sets achieved $\approx$$95\%$ size-stratified conditional coverage ($\alpha = 0.05$), as shown in \autoref{tab:cond-cvg}. This implies that, on average, there is a $95\%$ probability that an image's true label class falls within the prediction set, conditioned on all set sizes.

\begin{table}[ht]
\caption[]{Coverage achieved by RAPS across various set size ranges with error rate $\boldsymbol{\alpha=}\mathbf{0.05}$ for in-distribution stimuli in the \textit{conformal hold-out set}.}
\label{tab:cond-cvg}
\begin{tabular}{ccc}
\toprule
\multirow{2}{*}{Size} & \multicolumn{2}{c}{$\alpha = 0.05$} \\ \cline{2-3} 
 & Count & Coverage \\
\midrule
0 to 1 & 13,967 & 0.961 \\
2 to 15 & 6,849 & 0.946 \\
16 to 30 & 1,583 & 0.978 \\
31 to 45 & 874 & 0.975 \\
46 to 60 & 548 & 0.974 \\
101 to 1,000 & 1,179 & 0.974 \\
\bottomrule
\end{tabular}
\end{table}

\paragraph{\textbf{Generating OOD images}}
To compare the utility of different prediction displays when the classifier is poorly calibrated, we utilized OOD images, which would lower the accuracy of Top-$k$ predictions and increase the size of prediction sets.
To create OOD images, we systematically transformed in-distribution images from our \textit{conformal hold-out set} into OOD by applying \textit{covariate shifts}~\cite{quinonero2008dataset} through synthetic image corruption, as described by \citet{michaelis2019benchmarking}. Covariate shift, in the context of supervised learning where the joint distribution over images $X$ and labels $Y$ is $P(Y|X)P(X)$, refers to changes in $P(X)$, the feature probability, without affecting the classifier $P(Y|X)$.

After evaluating 15 types of corruption, we opted to use five corruptions that reduced WRN's Top-$1$ prediction accuracy the most. These corruptions were: (1) defocus blur, (2) snow, (3) frost, (4) zoom blur, and (5) glass blur, which are bolded in \autoref{tab:all-corruptions-top1-ranking} of Appendix~\ref{sec:appendix-methodol}. Given that the human visual system can be resilient to certain minor changes~\cite{michaelis2019benchmarking, hendrycks2019robustness}, we applied these corruptions to maximum severity for all images in the \textit{conformal hold-out set}, as defined by~\citet{michaelis2019benchmarking}.

\autoref{tab:all-corruptions-top1-ranking} of Appendix \ref{sec:appendix-methodol} indicates that our prediction sets achieve a $96\%$ coverage rate for in-distribution images. However, the coverage guarantee for the OOD samples was compromised, as the images used during calibration and testing are no longer exchangeable (i.i.d.), according to ~\citet{angelopoulos2020uncertainty}.

\paragraph{\textbf{Classifying stimuli by difficulty}} 
To compare the effects of different prediction displays on images of varying difficulty, we categorized images by difficulty using the cross-entropy loss, which calculates how well the predicted probability distribution aligns with the ground truth distribution. Formally in \autoref{eq:cross-entropy}, let $x$ be an input image; $p(x)$ denotes the true probability distribution of the image's label in one-hot encoding, while $q(x)$ is our classifier's predicted distribution. 

\begin{equation}\label{eq:cross-entropy}
H(p, q) = - \sum{p(x) \cdot \log(q(x))}    
\end{equation}

\autoref{fig:dist-indist-entropy} of Appendix \ref{sec:appendix-methodol} displays the distribution of cross-entropy loss of all in-distribution image stimuli, with the median presented as an orange dotted line. Due to WRN's high predictive power, this distribution was highly right-skewed, which created a problem: if we used this median threshold to categorize difficulty, we would include many images in the \texttt{easy} group that the model was both confident and accurate in predictions, so the derived sets embodied little uncertainty, synonymous with Top-$1$ prediction. We thus computed a new median threshold ($0.72$) after excluding images with a set size of one such that the difficulty threshold, shown as a red solid line, was defined based on images with non-deterministic sets. Images with cross-entropy loss falling below or on this threshold were classified as \texttt{easy}, while those above were deemed \texttt{hard}. 

Because the cross-entropy loss derived from the predictive distribution of OOD images can be misleading as a signal of task difficulty due to model misspecification~\cite{guo2017calibration,torralba2011unbiased}, we used the red median threshold from the in-distribution data to categorize the difficulty of OOD images.

\begin{table}[ht]
\caption[]{The median RAPS size of images binarized by difficulty within each \textit{conformal hold-out set}.}
\label{tab:set-size-threshold}
\begin{tabular}{ccccc}
\toprule
\multirow{2}{*}{Hold-out Set} & \multicolumn{2}{c}{w/ Size = 1} & \multicolumn{2}{c}{w/o Size = 1} \\ \cline{2-3} \cline{4-5}
 & Easy & Hard & Easy & Hard \\
\midrule
None & 1 & 15 & \textbf{5} & \textbf{19} \\
defocus blur & 2 & 59 & \textbf{8} & \textbf{60} \\
snow & 1 & 41 & \textbf{7} & \textbf{43} \\
frost & 2 & 46 & \textbf{8} & \textbf{50} \\
zoom blur & 2 & 44 & \textbf{10} & \textbf{47} \\
glass blur & 2 & 63 & \textbf{9} & \textbf{65}\\
\bottomrule
\end{tabular}
\end{table}

\paragraph{\textbf{Classifying stimuli by set size}}
Images within each difficulty group of the \textit{conformal hold-out set} induce varied prediction set sizes. For instance, while images classified as \texttt{easy} typically had a smaller set size, some \texttt{easy} instances required a larger set size to achieve the desired coverage. To explore how set size influences the efficacy of prediction sets, we categorized the stimuli within each difficulty group by their set size. After grouping images in each conformal hold-out set by \texttt{easy} and \texttt{hard}, we further distinguished them based on \textit{relative} set size, as determined by RAPS, into \texttt{smaller} or \texttt{larger} categories, based on the median of the group. However, as shown in \autoref{tab:set-size-threshold}, the predictive power of WRN consistently demonstrated high confidence in its predictions for many \texttt{easy} images with a set size of one, which significantly lowered the median size threshold. To introduce more variance in set size within the \texttt{easy} categories, we calculated the median threshold after excluding images with a set size of one (i.e., these were directly assigned to the \texttt{smaller} group). Images with a set size smaller than or equal to this adjusted size median were classified as \texttt{smaller}; if the opposite is true, they were classified as \texttt{larger}. The specific thresholds used for each difficulty category within the \textit{conformal hold-out set} are outlined in \autoref{tab:set-size-threshold}, with the bold figures representing the applied thresholds.

\paragraph{\textbf{Sampling task images based on defined categories}}\label{sec:generate-stimuli}
Upon categorizing all images in each \textit{conformal hold-out set} (i.e., one in-distribution and five OOD, totaling 6 $\times$ 25,000 images) by difficulty (\texttt{easy} and \texttt{hard}) and size (\texttt{smaller} and \texttt{larger}), our objective was to select a set of task images from which to sample image assignments for participants, so as to avoid results that overfit to a very small set of stimuli. Specifically, we targeted 80 task images from the 1,200 candidate images. Our goals were to ensure that: (1) the candidate images are representative of each group defined by in-distribution or OOD, difficulty, and size ($6 \times 2 \times 2$), and (2) task images avoid the known data anomalies in ILSVRC 2012~\cite{ekambaram2017finding,northcutt2021pervasive}.


\begin{table*}[htbp]
\centering
\caption[]{Prediction accuracy, counts, and set size of the selected 80 task stimuli grouped by our experimental manipulations.}
\label{tab:task-image-info}
\begin{tabular}{cccccccc}
\toprule
\multirow{2}{*}{\textbf{Shift}} & \multirow{2}{*}{\textbf{Difficulty}} & \multirow{2}{*}{\textbf{Size}} & \multirow{2}{*}{\textbf{Count}} & \multicolumn{3}{c}{\textbf{Coverage}} & \multirow{2}{*}{\textbf{Avg. Set Size}} \\ \cline{5-7}
                     &                       &         &    & \textbf{Top-1} & \textbf{Top-10} & \textbf{RAPS} &               \\ 
\midrule
\multirow{4}{*}{in}  & \multirow{2}{*}{easy} & smaller & 10 & 1     & 1      & 1    & 2.6  \\
                     &                       & larger  & 10 & 1     & 1      & 1    & 19.5 \\
                     & \multirow{2}{*}{hard} & smaller & 10 & 0.4   & 1      & 0.9  & 8.0  \\
                     &                       & larger  & 10 & 0.5   & 0.9    & 0.9  & 51.5 \\
\midrule
\multirow{4}{*}{out} & \multirow{2}{*}{easy} & smaller & 10 & 1     & 1      & 1    & \textbf{5}    \\
                     &                       & larger  & 10 & 1     & 1      & 1    & \textbf{28.1} \\
                     & \multirow{2}{*}{hard} & smaller & 10 & 0     & 0.6    & 0.9  & \textbf{30.1} \\
                     &                       & larger  & 10 & 0.1   & 0.2    & 0.9  & \textbf{90.8} \\ 
\bottomrule
\end{tabular}
\end{table*}

\begin{figure*}[ht]
  \centering
  \includegraphics[width=\linewidth]{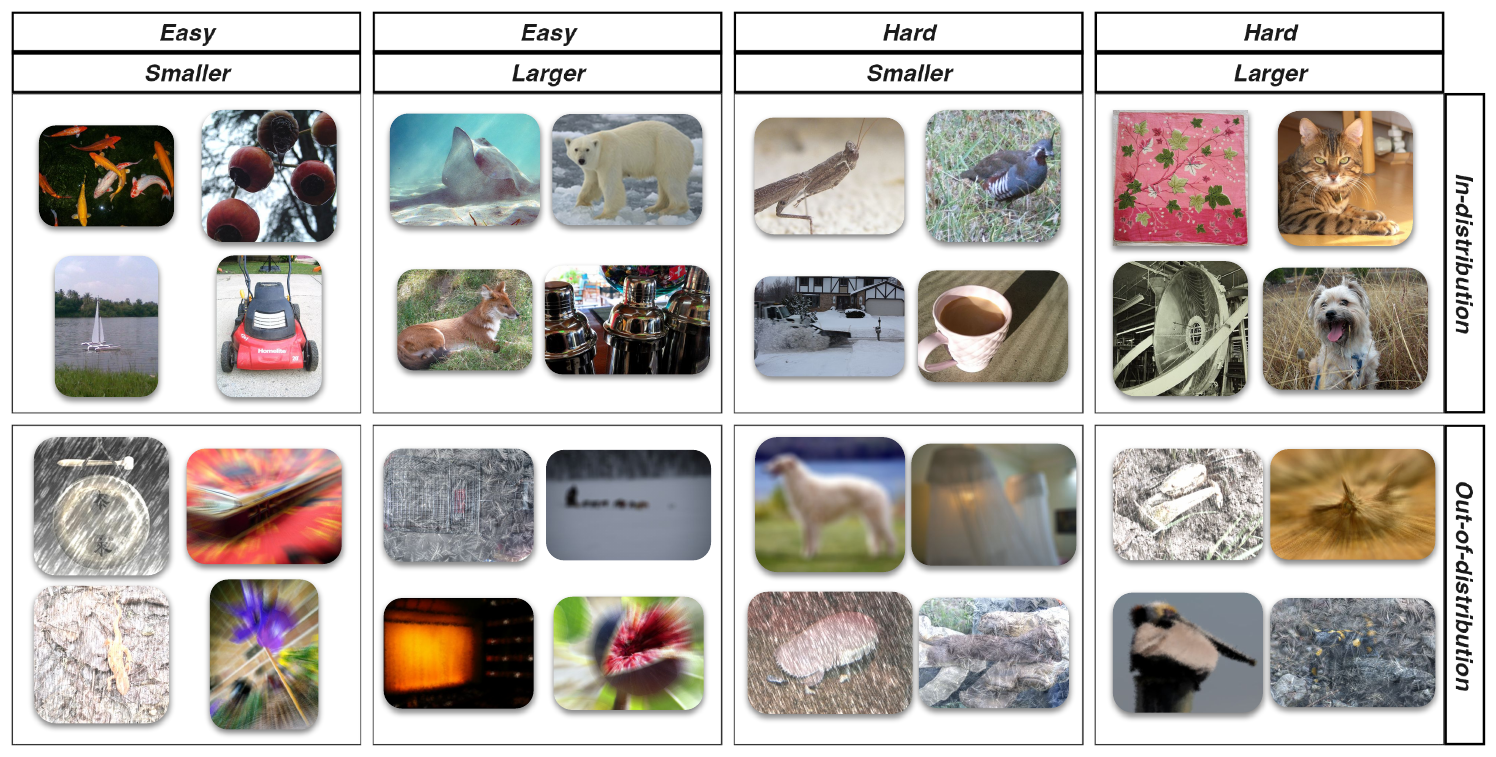}
  \caption[]{We present four example stimuli (from a total of 10) for each combination of in-distribution or OOD, difficulty, and size categories. The rows differentiate between in-distribution and OOD, while the columns vary by difficulty and set size.}
  \label{fig:stimuli-example}
  \Description{
  This figure displays a grid showcasing example stimuli participants viewed for labeling tasks. Rows differentiate in-distribution and OOD stimuli, while columns further categorize stimuli examples by difficulty and set size. 
  }
\end{figure*}

Sampling images from groups, such as \texttt{in-distribution}, \texttt{easy} (difficulty), and \texttt{smaller} (set size), was challenging due to the prevalence of images with a set size of one for which prediction sets become nearly identical to the Top-$1$ prediction. To avoid oversampling sets of size one, (1) we temporarily set aside images with a set size of one, (2) we took images with sets larger than one whose set size falls near the median size in that group (i.e., $45$\textsuperscript{th} and $55$\textsuperscript{th} percentiles), and (3) we performed a post-hoc sampling correction to include images with a set size of one in proportion to their original presence in each group, as shown in \autoref{tab:task-image-sample-correction} of Appendix~\ref{sec:appendix-methodol}.

Images from ILSVRC 2012 are known to be noisy, with numerous data anomalies. To create the set of 80 task images, we sampled 50 \textit{candidate} task images from each representative subset defined by in-distribution or OOD, difficulty, and size ($6 \times 2 \times 2 \times 50$) without replacement. We manually inspected the task images, starting from images with higher cross-entropy loss (\texttt{hard}), and omitted those with (1) an obviously wrong ground truth label, (2) a text overlay with the correct label, (3) no apparent focal object or many surrounding objects, (4) unusual dimensions, and (5) small multiples. Based on these rules, we selected 10 images from each of the 4 groups from the combination of $[\texttt{easy}, \texttt{hard}] \times [\texttt{smaller}, \texttt{larger}]$ that were in-distribution, and 2 images from each of the 4 groups with corruption. This process generated 80 task images in total, balanced by in-distribution or OOD, difficulty, and set size. 

\begin{figure*}[ht]
  \centering
  \includegraphics[width=0.9\linewidth]{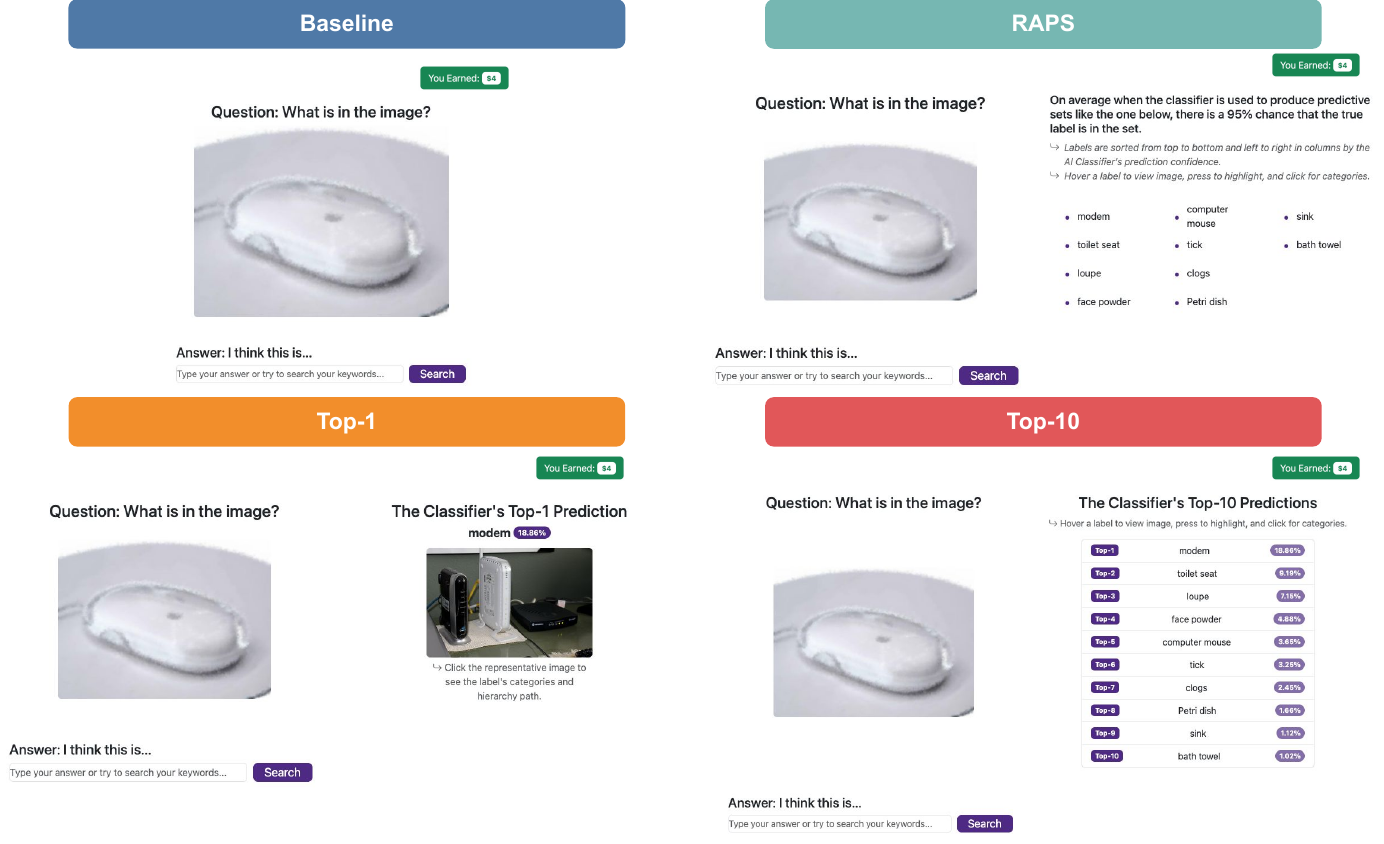}
  \caption[]{
  Screenshots of the interface participants used to complete the study by conditions (\textcolor[HTML]{4D79A7}{\textbf{\texttt{baseline}}}, \textcolor[HTML]{F28E2B}{\textbf{\texttt{Top-1}}}, \textcolor[HTML]{E15759}{\textbf{\texttt{Top-10}}}, \textcolor[HTML]{77B7B3}{\textbf{\texttt{RAPS}}}).
  }
  \label{fig:interface-display}
  \Description{
  This figure contains screenshots of the interface used by participants to complete the study, organized by conditions: baseline, Top-1, Top-10, and RAPS. Each screenshot of the interface illustrates the layout, adapted to accommodate different prediction displays with varying uncertainty quantification.
  }
\end{figure*}

\autoref{tab:task-image-info} presents the Top-$k$ accuracy, coverage, and average set size, grouped by our experimental factors, with example stimuli shown in \autoref{fig:stimuli-example}. Our stimuli exhibited the desired behavior, with the average prediction accuracy and set size varying in response to the manipulated difficulty and set size. While the stimuli we selected to achieve balance over our experimental manipulations resulted in close conformal coverage for OOD instances to the in-distribution case, anecdotally we found that OOD coverage tended to remain high ($\approx$$80\%$) over varying stimuli sets under our data generating model.

\subsubsection{Reward and Bonus}
Participants earned a base compensation of $\$4$ for completing the study and received a $\$0.25$ bonus for each correct label. With 16 trials, the maximum bonus was $\$4$, bringing the total potential earnings to $\$8$. We determined the task reward to achieve the local minimum wage based on the assumption that most participants would take longer than the average time we observed in a small pilot of the experiment. 

\subsection{Experimental Interface}
Our task interface, screenshots of which are presented in \autoref{fig:interface-display}, featured a two-column design: task images were on the left, and predictions were on the right, with a response field directly below the task image. In the baseline condition without predictions, the task image and the response field were centered. Participants could view their real-time earnings in the top-right corner of the interface.

To ensure that participants understood the meaning of each label, we selected label-representative images from the \textit{conformal calibration set}. To find suitable label-representatives, we manually examined all images by labels and prioritized selection for those with small cross-entropy loss (i.e., \texttt{easy}), applying the same rules used when selecting task images. The label-representative images we used are included in the Supplemental Material.

\subsubsection{Presenting Predictions}
For the three treatment conditions that featured predictions, the Top-$1$ condition presented the classifier's Top-1 prediction and its softmax score on top of a representative image of the predicted label. Top-$10$ predictions with softmax scores were ordered in a table, while the prediction set was presented in an $n$-by-$3$ matrix sorted column-wise by the classifier's prediction confidence. Participants could hover their mouse cursor over any predicted label to view a label-representative image. 

\begin{figure*}[ht]
  \centering
  \includegraphics[width=.8\linewidth]{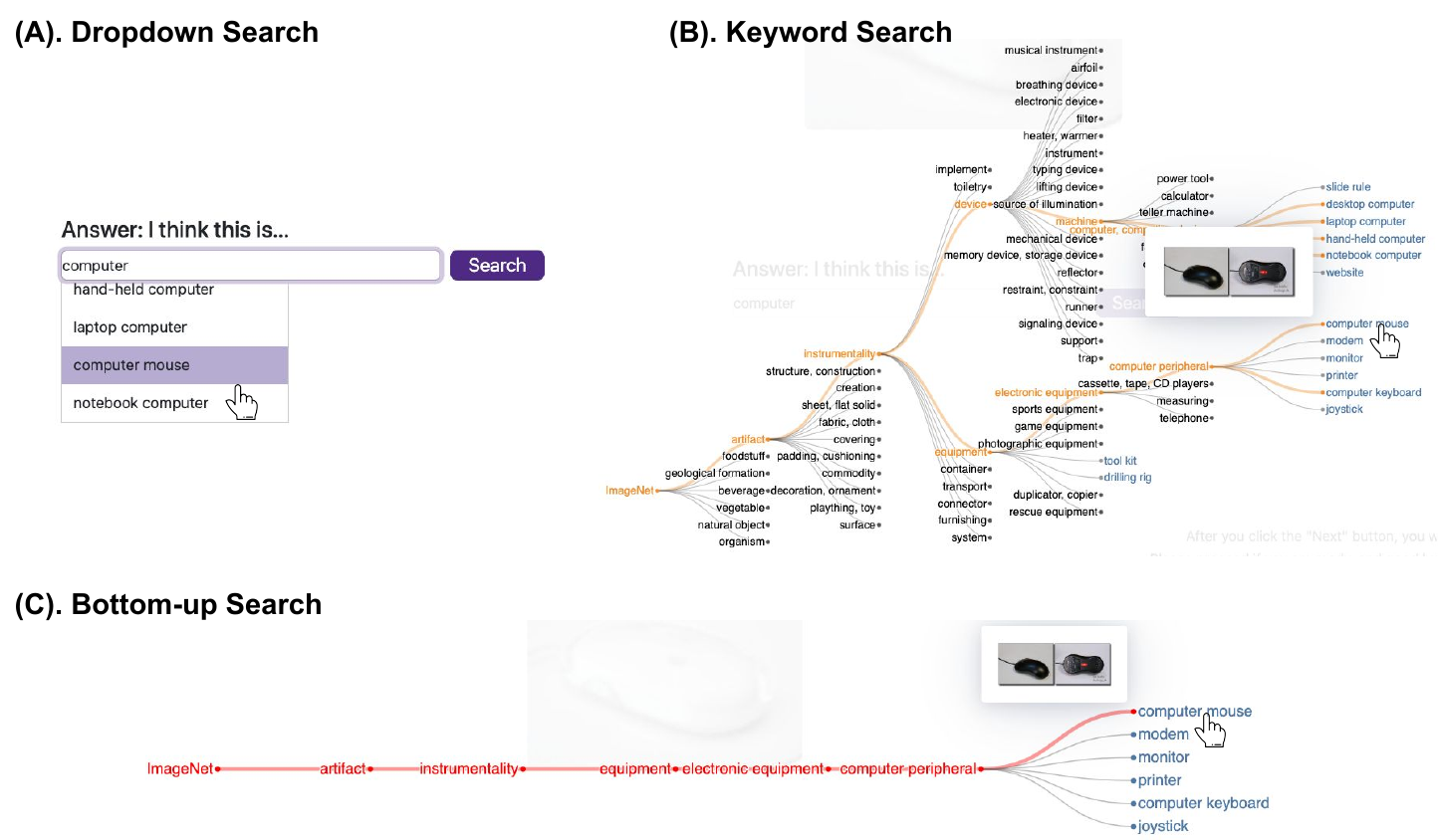}
  \caption[]{
  Participants are provided with three search options to find their preferred choice.
  (A) \textit{Dropdown search}: As participants type in the response field, a dropdown menu appears, exemplified by the entry ``computer''.
  (B) \textit{Keyword search}: Participants can search their typed keywords in the WordNet hierarchy by clicking the ``Search'' button, which opens an overlay displaying a hierarchy network with the relevant network components highlighted. We provide an example of the rendered hierarchy network by searching for ``computer''; 
  (C) \textit{Bottom-up search}: By clicking on a predicted label, such as ``computer mouse'', participants can see its path from the root node with categories on the path that can be clicked to expand for further exploration.
  Additionally, while exploring the hierarchy network, participants can hover over any leaf node to see label-representative images, as shown in (B) and (C).
  }
  \label{fig:search-tools}
  \Description{
    This figure contains screenshots to demonstrate the search features of the interface provided to participants to assist with their labeling tasks. When performing a dropdown search, a dropdown menu will appear below the input field, which shows potential labels that match what the participants typed. Keyword search enables participants to search for labels that match their typed keywords by clicking a ``Search" button. Bottom-up search allows participants to explore the label space by clicking on a predicted label. Additionally, participants can hover over any leaf node in the hierarchy network to access label-representative images.  
    }
\end{figure*}

\subsubsection{Label Space Search}
To label task images, participants need to be aware of the label space. Because ILSVRC 2012 has a large label space of 1,000, cumbersome to present in a dropdown or checkbox format, we organized labels using a version of the WordNet hierarchy, which we truncated slightly to avoid label overlap. For example, ``automobile" and ``truck" are both parent nodes for ``minivan" and since ``truck" is also a type of ``automobile", we retained ``minivan" in the ``automobile" category. We include this tree in the Supplemental Material. 

If participants disagreed with the predictions or when no prediction was available (baseline), our interface enabled them to perform three types of search. As illustrated in \autoref{fig:search-tools}, they were (A) dropdown search, (B) keyword search, and (C) bottom-up search. In the dropdown search, a dropdown menu would appear as participants began typing into the response field, suggesting leaf labels that matched their input. Alternatively, in the keyword search, participants could type and then click a search button to search the hierarchy for specific keywords. Doing so would open an overlay window showing the label-space hierarchy network, where leaf nodes (i.e., denoting valid labels) would appear in a different color from the parent nodes. Clicking on a leaf node would populate the label in the response field. To ease navigation, the hierarchy network was collapsed by default except for the parent nodes (i.e., categories) and leaf nodes (i.e., labels) containing the search keywords with edges connecting them highlighted in orange. All other parent nodes could also be clicked to expand.

Participants could also perform a bottom-up search starting from the prediction display. Clicking on a predicted label would open the search overlay. Instead of showing the entire hierarchy network, the overlay presented the path from the root node to the leaf node corresponding to the predicted label that the participant clicked on. Participants could explore adjacent labels under the same parent category or explore other parent nodes on the path toward the root.

When exploring the label space hierarchy, participants could hover over any leaf node to display its label-representative image. We illustrate these search actions through video demonstrations included in the Supplemental Material.

\subsection{Experimental Procedure}\label{exp-procedure}
Participants were directed to our study interface from Prolific. On the welcome page, they received a brief description of the study's purpose and the estimated time required for completion. Participants were instructed to complete the study in a single session using Google Chrome on a large-screen device. If participants agreed to the terms by clicking the ``Start'' button, they would be randomly assigned to a prediction display condition with task images sampled from each corresponding group, shuffled in a randomized order.

Participants then reviewed two instruction pages. The first introduced the interface layout and detailed the bonus mechanism. The second contained several short videos demonstrating how to explore predictions (except in the baseline condition) and navigate the label space. Participants had to watch all provided instructional videos before proceeding to an example page where they could practice using the interface. Then, they completed 16 rounds of tasks. Upon finishing the 16 trials, participants were asked to state their willingness-to-pay to have access to the same style of prediction display they used in the experiment on a scale from $0$ to $8$ USD (i.e., the range of possible earnings from the experiment) if they were invited to undertake another 16 rounds of tasks with new images generated similarly to those they had experienced, for further reward and potential bonus. On the same page, participants could optionally describe their strategy to identify the correct label for each image, with or without the aid of predictions. 

\subsection{Participants}
We recruited 600 participants from Prolific, divided equally among the four prediction displays, ensuring a gender-balanced sample. We set additional screening criteria so that all of our participants were (1) based in the USA, (2) had no vision impairment, (3) had no colorblindness, (4) spoke English as the first language, and (5) aged between 18 and 65. This led to 150 participants in each condition representing different prediction displays. We determined the target sample size using a non-parametric simulation of the experiment bootstrapped from a pilot study of 30 participants. By grouping trial-level responses according to all manipulated factors and bootstrapping these responses with replacement, we simulated trial-level outcomes. We identified the target sample size (600) that resulted in $95\%$ confidence intervals on accuracy with widths less than $10\%$. 

\subsection{Analysis Method}\label{sec:analysis-method}
We pre-registered an analysis plan focusing on three response variables: (1) accuracy, (2) shortest path length, and (3) willingness-to-pay, and followed our pre-registered plan in all analyses reported below unless otherwise specified. We define accuracy (ACC) as binary, indicating whether a participant selected the correct label for the task. The shortest path (SP) length is a continuous variable representing the number of hops from the leaf node a participant selects to the leaf node of the ground truth label. SP length is a measure widely used to quantify semantic similarity~\cite{resnik1995using, budanitsky2006evaluating} between categories in a taxonomy. Although highly correlated with accuracy, this measure provides further information on how ``incorrectness'' varied by our experimental manipulations. Willingness-to-pay (WTP) ranges from $\$0$ to $\$8$, measuring how much they will pay to access the display.

We pre-registered Bayesian Linear Mixed-effect models for accuracy and SP length. Our model specifications are presented in Model \ref{mod-acc} and Model \ref{mod-sp}. In this context, \textit{trial} is numeric, indicating the trial number, whereas \textit{condition} (4 levels: baseline, Top-$1$, Top-$10$, and prediction set), \textit{shift} (2 levels: in-distribution and OOD), \textit{difficulty} (2 levels: easy and hard), and \textit{size} (2 levels: smaller and larger) are factors corresponding to our experimental manipulations. The term \textit{ID} represents each participant's unique Prolific ID.

\begin{imageonly}
\begin{cpModel}
    \caption[]{Accuracy (ACC)}\label{mod-acc}
    \textit{acc} $\sim$ Bernoulli($p$) \label{lst:line:acc-likelihood} \;
    logit($p$)=\textit{condition} $\ast$ \textit{shift} $\ast$ \textit{difficulty} $\ast$ \textit{size} $\ast$ \textit{trial} + (\textit{trial}$|$\textit{ID}) \label{lst:line:acc-linear-model}
\end{cpModel}
\end{imageonly}

For the accuracy model stated in Model \ref{mod-acc}, we use a Bernoulli distribution for the likelihood shown in Line~\ref{lst:line:acc-likelihood}, parameterized by $p$, which denotes the probability of a correct label. Line \ref{lst:line:acc-linear-model} presents the hierarchical logistic regression model. In this specification, we model the logit of $p$ through interactions among all fixed effects and incorporate random intercepts and slopes for trial, grouped by participants' unique \textit{ID}. This approach allows us to account for participants' baseline performance differences and variations in rates of change across conditions and trials.

\begin{imageonly}
\begin{cpModel}
    \caption[]{Shortest Path (SP) Length}\label{mod-sp}
    \textit{sp} $\sim$ ZeroInflatedNegBinomial($\mu$, $\theta$, $\psi$) \label{lst:line:sp-likelihood} \;
    log($\mu$)=\textit{condition} $\ast$ \textit{shift} $\ast$ \textit{difficulty} $\ast$ \textit{size} $\ast$ \textit{trial} + (\textit{trial}|\textit{ID}) \label{lst:line:sp-linear-model} \;
    logit($\psi$)=\textit{condition} $\ast$ \textit{shift} $\ast$ \textit{difficulty} $\ast$ \textit{size} $\ast$ \textit{trial} \label{lst:line:sp-zero-inflation}
\end{cpModel}
\end{imageonly}

For the SP length model stated in Model \ref{mod-sp}, we employ a Zero-inflated Negative Binomial (ZINB) model as the likelihood, parameterized by $\mu$ (expected SP length), $\theta$ (negative binomial dispersion), and $\psi$ (zero-inflation probability), as shown in Line \ref{lst:line:sp-likelihood}. The expected path length between the participants' responses and the truth label is modeled by the negative binomial component in Line~\ref{lst:line:sp-linear-model}. We apply a log-link function to $\mu$, modeled by all fixed effects along with their interactions with random intercepts and slopes for trials grouped by participant ID. We do not explicitly model the dispersion parameter for the NB component. However, we placed an exponential prior with a rate of $0.1$ to moderate the degree of dispersion in our model. Given that our observed data had a maximum SP length of 27, this prior acts as a regularizing component, ensuring our model captures the dispersion of the observed data without predicting implausibly large SP lengths. Line~\ref{lst:line:sp-zero-inflation} shows the zero-inflation component that models the proportion of zero path length (i.e., correct answer) not captured by the NB components. We apply a logit link function to $\psi$ and model it by all fixed effects along with their interactions.

We deviated from our pre-registered model specifications only in setting the priors. We initially aimed to set weakly informative priors, but several were too narrow to allow model convergence. We therefore widened the priors to be even less informative. Full details are available in the Supplemental Material.

We followed a standard Bayesian workflow~\cite{gelman2020bayesian} to check if both models fit. Detailed model diagnostics are included in the Supplemental Material. We use the models to generate predictions of the respective response variable based on 1,000 draws from the posterior distribution of the models' fixed effects for each unique combination of our experimental manipulations while holding the trial effect constant at the mean. Our results present the expected predictions as point estimates, with uncertainty expressed through the $95\%$ highest posterior density interval (HDI).

We pre-registered an analysis plan to assess participants' reported WTP, relative to the aggregate observed value of each prediction display. Unlike traditional Likert-style questions, WTP allows us to define an objective benchmark post-hoc for comparison to participants' responses.

Specifically, we first calculate the \textit{expected bonus} participants gained in each condition by multiplying the expected accuracy rates by the maximum earnable bonus of $\$4$. We then calculate the \textit{expected bonus difference} with access to each prediction display relative to the baseline by subtracting the \textit{expected bonus} of the baseline from the \textit{expected bonus} of each treatment condition. We finally evaluate the discrepancy between this \textit{expected bonus difference} and the corresponding WTPs reported by participants who used each prediction display by constructing a ``\textit{willingness-to-overpay}'' metric. This measure reflects the difference between the additional bonus one is expected to earn from having access to a prediction display over the baseline and the actual difference in the average bonus participants received from accessing a prediction display relative to the baseline. We use bootstrapping to derive 95\% confidence intervals with the median as a point estimate.

\subsection{Qualitative Analysis of Strategies}
We performed a qualitative analysis and developed a bottom-up open coding scheme using the strategies provided by the participants. Two coders worked in collaboration, each coding half of the strategies. After coding, any ambiguous strategies were discussed until mutual agreements were reached. Among 585 valid strategies, we excluded 41 uninformative ones (Code: 1) with responses such as ``I have no idea, trying my hardest''. For the remaining strategies, we identified 17 ways in which the prediction displays and the search features were used, as illustrated in \autoref{tab:code-summary}.

\begin{table}[ht]
\centering
\caption[]{Summary of participants' strategy categories derived from open coding.}
\label{tab:code-summary}
\begin{tabularx}{\columnwidth}{@{}cX@{}}
\toprule
\textbf{Code}  & \textbf{Category}\\ 
\midrule
2--6  & How do participants use different prediction displays? \\
7--10 & Do participants find predictions to be trustworthy?    \\
11--15 & How do participants use the search feature to find a preferred label?         \\
16--18 & How do participants narrow down their choices and identify a preferred label? \\ 
\bottomrule
\end{tabularx}
\end{table}

We provide detailed code descriptions in \autoref{tab:open-codes-description} of Appendix~\ref{sec:appendix-methodol}, and generate a spreadsheet of codes and quotes to summarize the diversity and nuances in participants' strategies, which are included in the Supplemental Material.
\section{Results}\label{sec:res}

\subsection{Data Preliminaries}
\begin{figure}[ht]
  \centering
  \includegraphics[width=\linewidth]{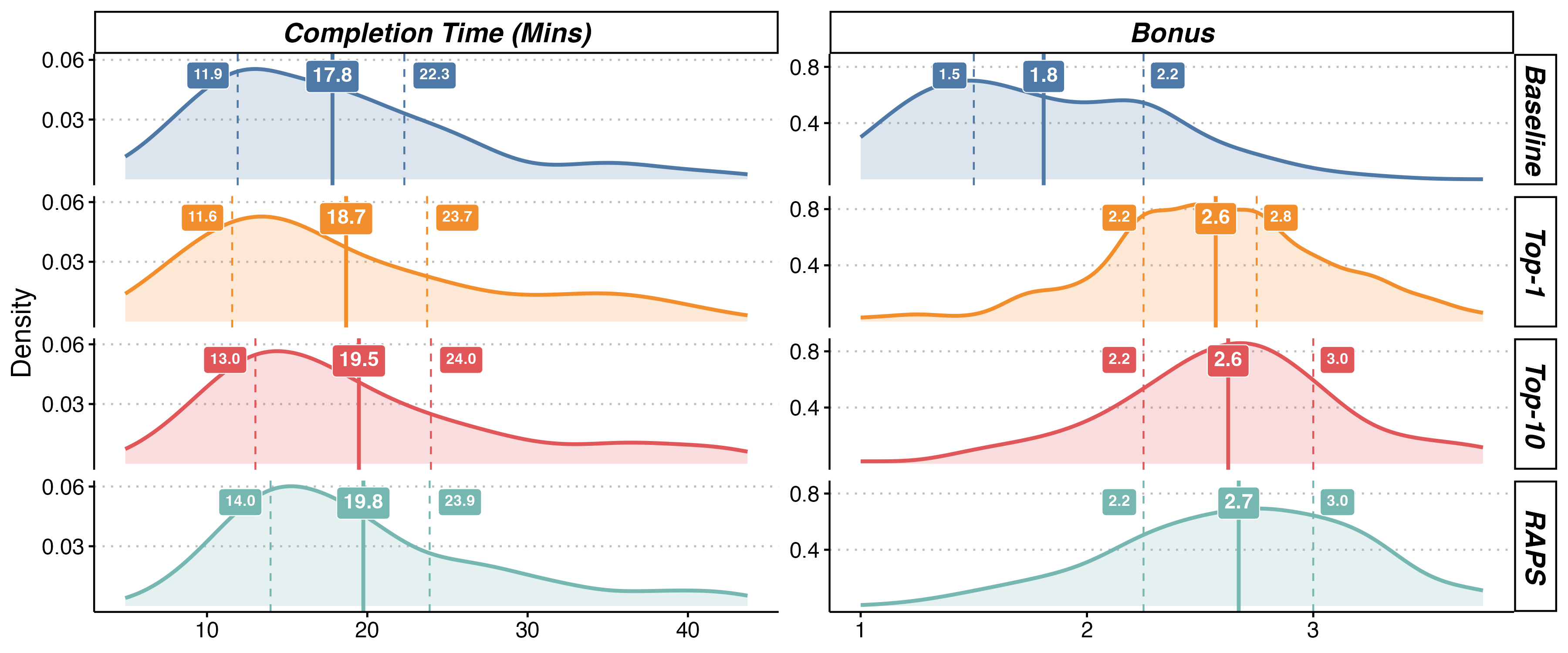}
  \caption[]{
  Distribution of participants' study completion time (minutes) and bonus earned by conditions with a solid vertical line showing average and dotted vertical lines showing the interquartile range.
  }
  \label{fig:results-overview}
  \Description{
  This figure presents the distributions of participants' study completion time and the amount of bonus earned by treatment conditions. Each distribution has a solid vertical line showing the average and two dotted vertical lines showing the interquartile range.
  }
\end{figure}

We received 599 valid responses after excluding one participant from the \textcolor[HTML]{4D79A7}{\textbf{\texttt{baseline}}} condition according to our pre-registered exclusion rule. As demonstrated in \autoref{fig:results-overview} left, participants spent, on average, $\approx$$20$ minutes completing the 16 image labeling tasks. Participants in the \textcolor[HTML]{E15759}{\textbf{\texttt{Top-10}}} and \textcolor[HTML]{77B7B3}{\textbf{\texttt{RAPS}}} conditions took slightly longer to complete tasks than those in the \textcolor[HTML]{4D79A7}{\textbf{\texttt{baseline}}} and \textcolor[HTML]{F28E2B}{\textbf{\texttt{Top-1}}} conditions.

From the distribution of earned bonuses shown in \autoref{fig:results-overview} right, there is a clear difference indicating that access to the prediction displays helped participants achieve more correct answers and hence a higher bonus reward than those without access. On average, \textcolor[HTML]{4D79A7}{\textbf{\texttt{baseline}}} participants provided seven correct labels, which is lower than those who used prediction displays (10 out of 16 correct).

\subsection{Accuracy}

\subsubsection{Overview}
\begin{figure*}[ht]
  \centering
  \includegraphics[width=\linewidth]{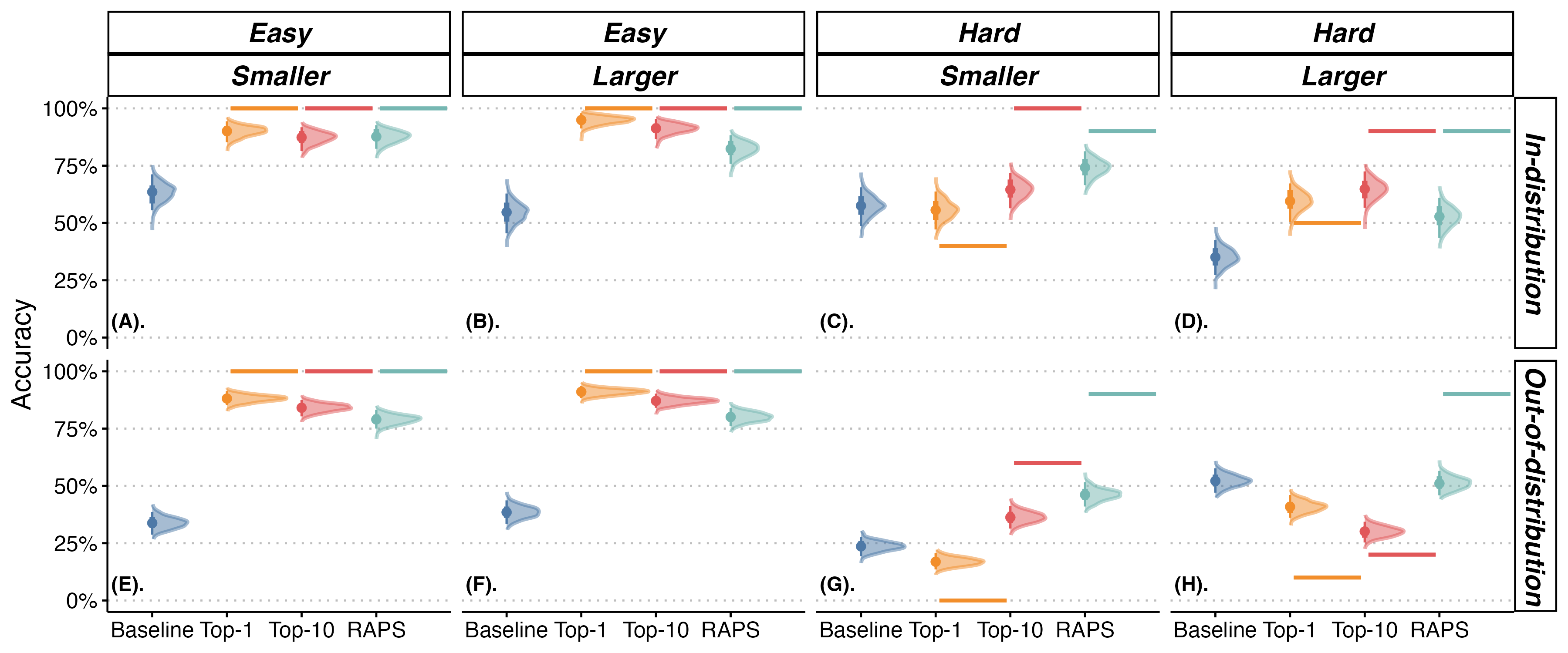}
  \caption[]{
  Median expected labeling accuracy for in-distribution and OOD task images grouped by difficulty (\texttt{easy} and \texttt{hard}) and set size (\texttt{smaller} and \texttt{larger}) predicted by the accuracy model (Model \ref{mod-acc}) with uncertainty expressed as 95\% HDIs. Solid horizontal lines show the prediction accuracy achieved by that display (i.e., the probability that the displayed predictions included the true label) using our base classifier (i.e., WRN). Note: for easy in-distribution stimuli, the average set sizes are 2.6 (\texttt{smaller}) and 19.5 (\texttt{larger}); for hard, they are 8 (\texttt{smaller}) and 51.1 (\texttt{larger}). For easy OOD stimuli, the averages are 5 (\texttt{smaller}) and 28.1 (\texttt{larger}); for hard OOD, they are 30.1 (\texttt{smaller}) and 90.8 (\texttt{larger}).
  }
  \label{fig:acc-results}
  \Description{
  This figure displays a grid presenting the median expected labeling accuracy predicted by our accuracy model with uncertainty expressed as 95\% HDIs. Rows differentiate in-distribution and OOD results, while columns separate results by difficulty and set size. Solid horizontal lines for each stimuli group illustrate the prediction accuracy achieved by that display (i.e., the probability that the displayed predictions included the true label) using our base classifier (i.e., WRN). 
  }
\end{figure*}

In \autoref{fig:acc-results}, we present the expected predictions by our accuracy model (Model \ref{mod-acc}) for each type of image stimuli, with uncertainty quantified as $95\%$ HDIs. At a high level, we find that \textcolor[HTML]{77B7B3}{\textbf{\texttt{RAPS}}} does not improve participants' labeling accuracy for \texttt{easy} tasks but is more useful for \texttt{hard} tasks, especially for \texttt{OOD} images. While a \texttt{smaller} \textcolor[HTML]{77B7B3}{\textbf{\texttt{RAPS}}} can be more useful for in-distribution images, size does not appear to affect accuracy much when images are \texttt{OOD}. Because \texttt{smaller} and \texttt{larger} set sizes depend on whether instances are \texttt{in-distribution} or \texttt{OOD} in our data-generating model (Section~\ref{sec:dgm}), we provide the average set size for size groups to contextualize the results below.

\subsubsection{In-distribution}

\paragraph{\textbf{For both \texttt{easy} and \texttt{hard} images that are \texttt{in-distribution}, a \texttt{larger} set size may decrease participants' labeling accuracy}}
This effect is most prominent when images are classified as \texttt{hard} in \autoref{fig:acc-results}, (C) and (D). When using \textcolor[HTML]{77B7B3}{\textbf{\texttt{RAPS}}} to label \texttt{hard} images, participants' accuracy is higher when the set size is \texttt{smaller} ($74.2\%$; HDI: $[66.5\%, 81.3\%]$; average size: $2.6$), and lower when set size is \texttt{larger} ($52.8\%$; HDI: $[43.5\%, 60.9\%]$; average size: $19.5$). We speculate that the reason is that a \texttt{larger} set size increases participants' cognitive load as they need to traverse through the set and evaluate all labels. \autoref{fig:acc-results} (C) shows that expected accuracy of \textcolor[HTML]{77B7B3}{\textbf{\texttt{RAPS}}} participants is roughly $8$\% higher than that of the \textcolor[HTML]{E15759}{\textbf{\texttt{Top-10}}} condition ($64.5\%$; HDI: $[56.4\%, 71.7\%]$) which is roughly $10$\% higher than the \textcolor[HTML]{F28E2B}{\textbf{\texttt{Top-1}}} condition ($55.6\%$; HDI: $[47.2\%, 63.7\%]$), though HDIs overlap.

\paragraph{\textbf{When the true label is in the prediction display, participants do not always choose it.}} This is particularly evident for \texttt{in-distribution} images categorized as \texttt{easy} (cf.~\autoref{fig:acc-results}, A and B). Here, the participants' labeling accuracy tends to be lower than the prediction display accuracy (i.e., the percentage of the time the prediction display included the true label), depicted as solid horizontal lines. However, when labeling \texttt{hard} images, participants in the \textcolor[HTML]{F28E2B}{\textbf{\texttt{Top-1}}} condition can improve upon the model's prediction. As suggested by our open codes discussed in Section~\ref{sec:strategies}, we find that when the \textcolor[HTML]{F28E2B}{\textbf{\texttt{Top-1}}} prediction was obviously inaccurate, participants were more likely to rely on their own judgment and used the search tool to identify the correct answer. The presence of a greater number of labels in the \textcolor[HTML]{E15759}{\textbf{\texttt{Top-10}}} and \textcolor[HTML]{77B7B3}{\textbf{\texttt{RAPS}}} conditions may have complicated the labeling process due to an increased cognitive load. This is exemplified by the observed trends in these conditions: among incorrect responses, participants in the \textcolor[HTML]{77B7B3}{\textbf{\texttt{RAPS}}} condition missed the correct label from the prediction set $60\%$ of the time, whereas those in the \textcolor[HTML]{E15759}{\textbf{\texttt{Top-10}}} condition did so $80\%$ of the time.

In summary, when task images are \texttt{in-distribution}, (1) labeling accuracy for participants in the \textcolor[HTML]{F28E2B}{\textbf{\texttt{Top-1}}} or \textcolor[HTML]{E15759}{\textbf{\texttt{Top-10}}} conditions is higher when the images are \texttt{easy} (cf. \autoref{fig:acc-results}, A and B); (2) the accuracy of \textcolor[HTML]{77B7B3}{\textbf{\texttt{RAPS}}} participants is negatively correlated with set size: a \texttt{larger} set size decreases accuracy, particularly for \texttt{hard} images (cf. \autoref{fig:acc-results}, C and D); and (3) \textcolor[HTML]{F28E2B}{\textbf{\texttt{Top-1}}} participants are more likely to rely on their own judgment and use the search tool to identify the correct answer when predictions are less likely to be correct. In contrast, the pattern of incorrect responses observed in the \textcolor[HTML]{E15759}{\textbf{\texttt{Top-10}}} and \textcolor[HTML]{77B7B3}{\textbf{\texttt{RAPS}}} conditions suggest that the presence of more predictions makes it harder for participants to discern between inaccurate and accurate predictions, even when the correct label is present.

\subsubsection{Out-of-distribution} \label{sec:result-acc-ood}

\paragraph{\textbf{When \texttt{OOD} images are \texttt{easy}, Top-\boldmath$k$ predictions yield higher accuracy than \textcolor[HTML]{77B7B3}{\textbf{\texttt{RAPS}}}}} 
Due to the similarity in HDIs across conditions for \texttt{easy} \texttt{OOD} images (cf. \autoref{fig:acc-results}, E and F), we report expected accuracy marginalized over set sizes. We find \textcolor[HTML]{F28E2B}{\textbf{\texttt{Top-1}}} participants tend to achieve slightly higher accuracy in expectation ($89.6\%$; HPI: $[85.5\%, 93.4\%]$) than those using \textcolor[HTML]{E15759}{\textbf{\texttt{Top-10}}} ($85.7\%$; HDP: $[81\%, 89.6\%]$). Meanwhile, the \textcolor[HTML]{77B7B3}{\textbf{\texttt{RAPS}}} condition yields the lowest expected accuracy of $79.5\%$ (HPI: $[75.1\%, 83.6\%]$), despite an average set size of $5$ (as detailed in \autoref{tab:task-image-info}) that is smaller than the size of \textcolor[HTML]{E15759}{\textbf{\texttt{Top-10}}}, with identical chance that the prediction display contained the true label (i.e., colored horizontal lines in \autoref{fig:acc-results}, E and F). This divergence from our \texttt{in-distribution} finding, where \texttt{smaller} set sizes typically correlate with higher accuracy when predictions are reliable, suggests that additional factors may be at play in the \texttt{OOD} scenario. It is possible that the reduction in cognitive load afforded by having access to softmax scores in the \textcolor[HTML]{E15759}{\textbf{\texttt{Top-10}}} condition aided participants in finding the correct label. This observation partially reinforces our finding from the \texttt{easy} \texttt{in-distribution} results: when the model is well-calibrated (cf. \autoref{fig:acc-results}, E and F), viewing fewer predictions and having access to prediction confidence (i.e., as in our Top-$k$ conditions) can improve accuracy.

\begin{figure*}[ht]
  \centering
  \includegraphics[width=\linewidth]{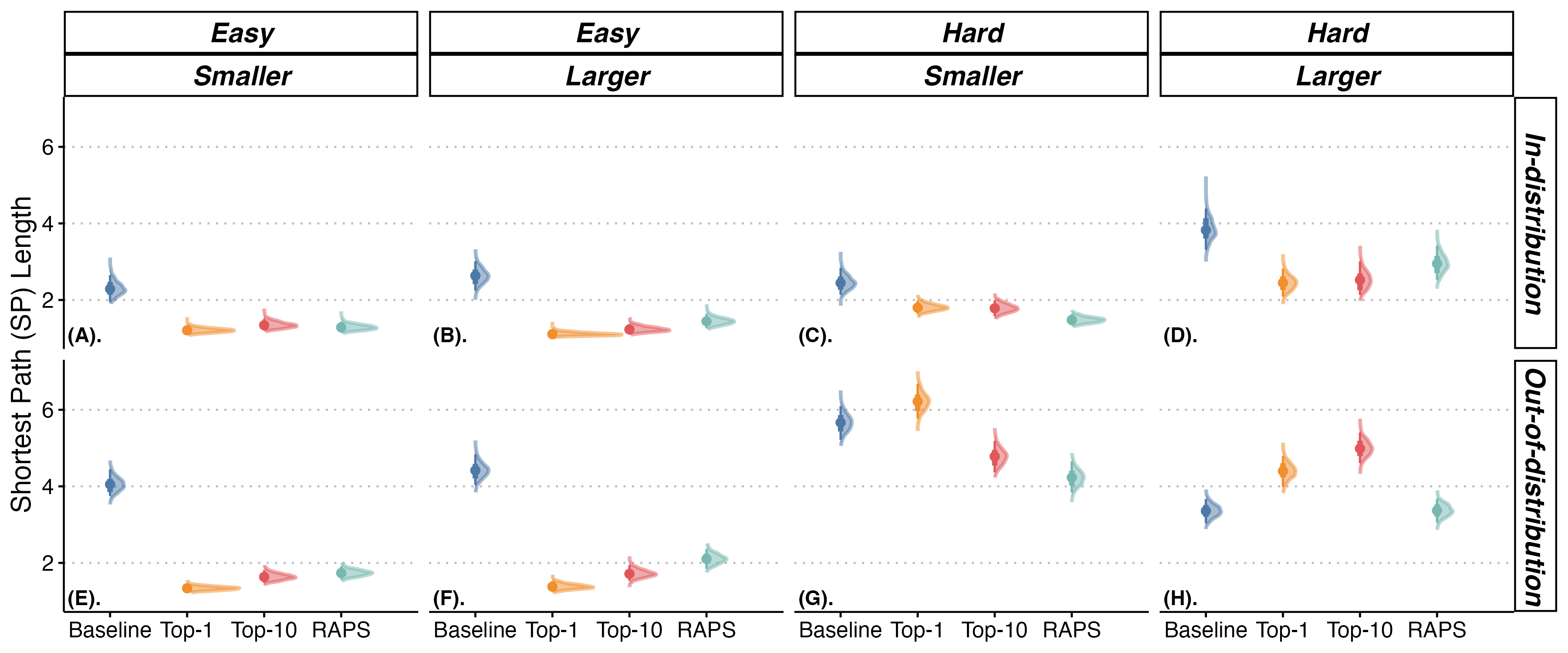}
  \caption[]{
  Median expected shortest path length quantifies the amount of label error for \texttt{in-distribution} and \texttt{OOD} task images grouped by difficulty (\texttt{easy} and \texttt{hard}) and set size (\texttt{smaller} and \texttt{larger}) predicted by the shortest path model (Model \ref{mod-sp}) with uncertainty expressed through 95\% HDIs. Note: for easy in-distribution stimuli, the average set sizes are 2.6 (\texttt{smaller}) and 19.5 (\texttt{larger}); for hard, they are 8 (\texttt{smaller}) and 51.1 (\texttt{larger}). For easy OOD stimuli, the averages are 5 (\texttt{smaller}) and 28.1 (\texttt{larger}); for hard OOD, they are 30.1 (\texttt{smaller}) and 90.8 (\texttt{larger}).
  }
  \label{fig:sp-results}
  \Description{
  This figure displays a grid presenting the median expected shortest path length predicted by our shortest path model with uncertainty expressed as 95\% HDIs. Rows differentiate in-distribution and OOD results, while columns separate results by difficulty and set size.
  }
\end{figure*}

\paragraph{\textbf{When \texttt{OOD} images are \texttt{hard}, \textcolor[HTML]{77B7B3}{\textbf{\texttt{RAPS}}} yield the highest labeling accuracy regardless of the set size}} 
For \texttt{hard} \texttt{OOD} images, \textcolor[HTML]{77B7B3}{\textbf{\texttt{RAPS}}} set sizes grow substantially larger, with \texttt{smaller} sets averaging $30$ instances and \texttt{larger} sets averaging $91$ instances. When \texttt{hard OOD} images had \texttt{smaller} set sizes (i.e., \autoref{fig:acc-results}, G), \textcolor[HTML]{77B7B3}{\textbf{\texttt{RAPS}}} participants outperform those in Top-$k$ conditions with a relatively high accuracy of $45.1\%$ (HPI: $[41\%, 51.6\%]$), followed by the \textcolor[HTML]{E15759}{\textbf{\texttt{Top-10}}} condition ($36.3\%$; HPI: $[31.4\%, 41.3\%]$) and \textcolor[HTML]{F28E2B}{\textbf{\texttt{Top-1}}} condition ($16.9\%$; HPI: $[13.5\%, 20.7\%]$). When the set size is \texttt{larger} (i.e., \autoref{fig:acc-results}, H), we identify a consistent pattern that \textcolor[HTML]{77B7B3}{\textbf{\texttt{RAPS}}} participants ($51.1\%$, HPI: $[45.9\%, 56.5\%]$) outperform those in the Top-$k$ conditions. Nevertheless, the accuracy of \textcolor[HTML]{77B7B3}{\textbf{\texttt{RAPS}}} participants is still lower than the average coverage rate (cf. \autoref{fig:acc-results}, G and F, with \textcolor[HTML]{77B7B3}{\textbf{\texttt{green}}} solid horizontal lines), implying that even when the sets include the true label, it is still challenging for participants to identify it.

\paragraph{\textbf{When \texttt{OOD} images are \texttt{hard}, participants relying on their own judgment may achieve better accuracy than those relying on poorly-calibrated Top-\boldmath$k$ predictions}}
For \texttt{hard} \texttt{OOD} images with \texttt{smaller} average set size (i.e., \autoref{fig:acc-results}, G), we find that participants in the \textcolor[HTML]{4D79A7}{\textbf{\texttt{baseline}}} condition who did not have access to any predictions have a higher accuracy of $23.6\%$ (HPI: $[19.3\%, 27.6\%]$) than those in the \textcolor[HTML]{F28E2B}{\textbf{\texttt{Top-1}}} condition ($16.9\%$; HPI: $[13.5\%, 20.7\%]$). For \texttt{hard} \texttt{OOD} images with a \texttt{larger} average set size (i.e., \autoref{fig:acc-results}, H), the accuracy of using \textcolor[HTML]{77B7B3}{\textbf{\texttt{RAPS}}} ($51.1\%$, HPI: $[45.9\%, 56.5\%]$) versus the \textcolor[HTML]{4D79A7}{\textbf{\texttt{baseline}}} condition ($52.2\%$; HPI: $[47\%, 57.7\%]$) is similar, and participants who used \textcolor[HTML]{F28E2B}{\textbf{\texttt{Top-1}}} prediction are more accurate ($40.9\%$; HPI: $[36.1\%, 46\%]$) than those who used \textcolor[HTML]{E15759}{\textbf{\texttt{Top-10}}} predictions ($30.1\%$; HPI: $[25.3\%, 34.4\%]$). 
Our results suggest that Top-$k$ predictions from a poorly calibrated model (cf. \autoref{fig:acc-results}, G and H) can negatively influence human judgment relative to no model assistance. Although \textcolor[HTML]{77B7B3}{\textbf{\texttt{RAPS}}} participants have relatively higher accuracy than those in the Top-$k$ conditions, their accuracy is almost identical to those in \textcolor[HTML]{4D79A7}{\textbf{\texttt{baseline}}}, suggesting it might be beneficial to withhold predictions for \texttt{hard} \texttt{OOD} stimuli that require a \texttt{larger} set size to achieve the desired coverage, at least in an OOD setting like ours, where the perturbations decreased both human and AI performance.

In summary, (1) when \texttt{OOD} images are \texttt{easy} and the predictions and softmax scores well-calibrated (cf. \autoref{fig:acc-results}, E and F), Top-$k$ predictions tend to result in higher labeling accuracy; (2) when \texttt{OOD} images are \texttt{hard} and the model predictions more error-prone (cf. \autoref{fig:acc-results}, G and H), \textcolor[HTML]{77B7B3}{\textbf{\texttt{RAPS}}} participants can outperform their Top-$k$ counterparts. Unlike in the in-distribution case, increasing the size of the prediction set does not degrade accuracy; and (3) when \texttt{OOD} images are \texttt{hard} and the model predictions are more error-prone, participants who rely on their own judgment and the label space search tool may be more likely to identify the correct answer. The reasons behind the performance advantage of the participants in the \textcolor[HTML]{4D79A7}{\textbf{\texttt{baseline}}} condition when labeling \texttt{hard} \texttt{OOD} images with \texttt{larger} set sizes remain ambiguous. It is possible that the accuracy decrease for participants in \textcolor[HTML]{E15759}{\textbf{\texttt{Top-10}}} condition when labeling \texttt{hard} \texttt{OOD} images that have \texttt{larger} set size is due to overreliance on inaccurate predictions, given that the average coverage for \textcolor[HTML]{E15759}{\textbf{\texttt{Top-10}}} predictions decreases from $60\%$ (\texttt{smaller}) to $20\%$ (\texttt{larger}), as shown in \autoref{tab:task-image-info}.

\subsection{Shortest Path Length}
The predictions of our SP length model (Model \ref{mod-sp}) are highly correlated with those of the accuracy model when comparing the results in \autoref{fig:acc-results} with those in \autoref{fig:sp-results}. We briefly highlight the main takeaways that complement our accuracy results, with more detailed descriptions in Appendix~\ref{sec:appendix-sp-results}.

\paragraph{\textbf{When OOD images are \texttt{hard}, a \texttt{larger} prediction set leads to submitted labels that are slightly closer to the correct label}}
Recall that in Section~\ref{sec:result-acc-ood}, we find when \texttt{OOD} images are classified as \texttt{hard}, \textcolor[HTML]{77B7B3}{\textbf{\texttt{RAPS}}} participants can outperform their Top-$k$ counterparts, but there is no difference in accuracy by set size (cf. \autoref{fig:acc-results}, G and H). However, from the SP length, we find that if the task image has a \texttt{larger} set size, \textcolor[HTML]{77B7B3}{\textbf{\texttt{RAPS}}} participants' chosen label is slightly closer to the correct answer in the label space hierarchy. As demonstrated in~\autoref{fig:sp-results} (G) and (H), the SP length for \texttt{hard} OOD images that have a \texttt{larger} set size is $3.4$ (HPI: $[3.1, 3.7]$, average size: $90.8$) and that of a \texttt{smaller} set size is 4.2 (HPI: $[3.8, 4.7]$; average size: $30.1$). It is difficult to say why this might be the case; it is possible, as supported by our open codes in Section~\ref{sec:strategies}, that when \texttt{OOD} stimuli are especially challenging to recognize, participants spend more time scrutinizing labels in the prediction set for clues, and that even when the probability that the prediction set contains the true label is the same, viewing more predictions helps cue their ability to identify the true label. The effect is quite small, whatever the cause.

\paragraph{\textbf{When predictions are generated by a well-calibrated model, having access to predictions can guide participants closer to the correct answer}} 
This result is most prominent for \texttt{hard} \texttt{in-distribution} images with \texttt{smaller} set size. From \autoref{fig:acc-results} (C), we see the \textcolor[HTML]{4D79A7}{\textbf{\texttt{baseline}}} and \textcolor[HTML]{F28E2B}{\textbf{\texttt{Top-1}}} prediction tend to yield similar accuracy, as HDIs closely overlap. However, in \autoref{fig:sp-results} (C), we find that participants who used \textcolor[HTML]{F28E2B}{\textbf{\texttt{Top-1}}} prediction can provide responses that are closer to the correct answer than those in the \textcolor[HTML]{4D79A7}{\textbf{\texttt{baseline}}}.

In summary, we find (1) for \texttt{hard} OOD stimuli in our setting, although the accuracy of \textcolor[HTML]{77B7B3}{\textbf{\texttt{RAPS}}} participants is similar, a \texttt{larger} set may guide responses closer to the ground truth (cf. \autoref{fig:sp-results}, G and H); and (2) when the model is well-calibrated, inaccurate Top-$k$ predictions might help participants in identifying the correct label. 

\subsection{Willingness-to-Pay}
We present the distributions of elicited willingness-to-pay (WTP) measures in \autoref{fig:wtp-dist} with summary statistics in \autoref{tab:wtp-descriptive} of Appendix~\ref{sec:appendix-wtp}. Elicited WTPs across conditions are quite similar, suggesting that the measure may be noisy. Participants are willing to pay slightly more for \textcolor[HTML]{E15759}{\textbf{\texttt{Top-10}}} ($\$1.84$; CI: $[1.58, 2.11]$) than \textcolor[HTML]{77B7B3}{\textbf{\texttt{RAPS}}} ($\$1.78$; CI: $[1.51, 2.07]$) and \textcolor[HTML]{F28E2B}{\textbf{\texttt{Top-1}}} ($\$1.7$; CI: $[1.46, 1.96]$).

Nonetheless, as detailed in Section~\ref{sec:analysis-method}, we quantify participants' perceived utility for each prediction display using a measure we devised, called ``\textit{willingness-to-overpay}''. This proxy focuses on economic valuation, offering direct monetary comparisons between the monetary value of a prediction display that participants perceive (i.e., WTP) against how much additional expected bonus participants gained from accessing each prediction display.

\begin{table}[h]
\caption[]{\textit{Expected Bonus Diff.} column shows the additional expected bonus participants gained when completing tasks using each prediction display relative to the \textcolor[HTML]{4D79A7}{\textbf{\texttt{baseline}}}.}
\label{tab:wtp-nominal}
\begin{tabular}{@{}cccc@{}}
\toprule
\textbf{Condition} & \textbf{Accuracy} & \begin{tabular}[c]{@{}c@{}}\textbf{Expected}\\ \textbf{Bonus}\end{tabular} & \begin{tabular}[c]{@{}c@{}}\textbf{Expected}\\ \textbf{Bonus Diff.}\end{tabular} \\ 
\midrule
\textcolor[HTML]{4D79A7}{\textbf{\texttt{baseline}}} & 0.41 & 1.64 & 0.00 \\
\textcolor[HTML]{F28E2B}{\textbf{\texttt{Top-1}}} & 0.63 & 2.52 & 0.86 \\
\textcolor[HTML]{E15759}{\textbf{\texttt{Top-10}}} & 0.64 & 2.56 & 0.88 \\
\textcolor[HTML]{77B7B3}{\textbf{\texttt{RAPS}}} & 0.66 & 2.64 & 0.98 \\
\bottomrule
\end{tabular}
\end{table}

\begin{figure}[ht]
  \centering
  \includegraphics[width=0.9\linewidth]{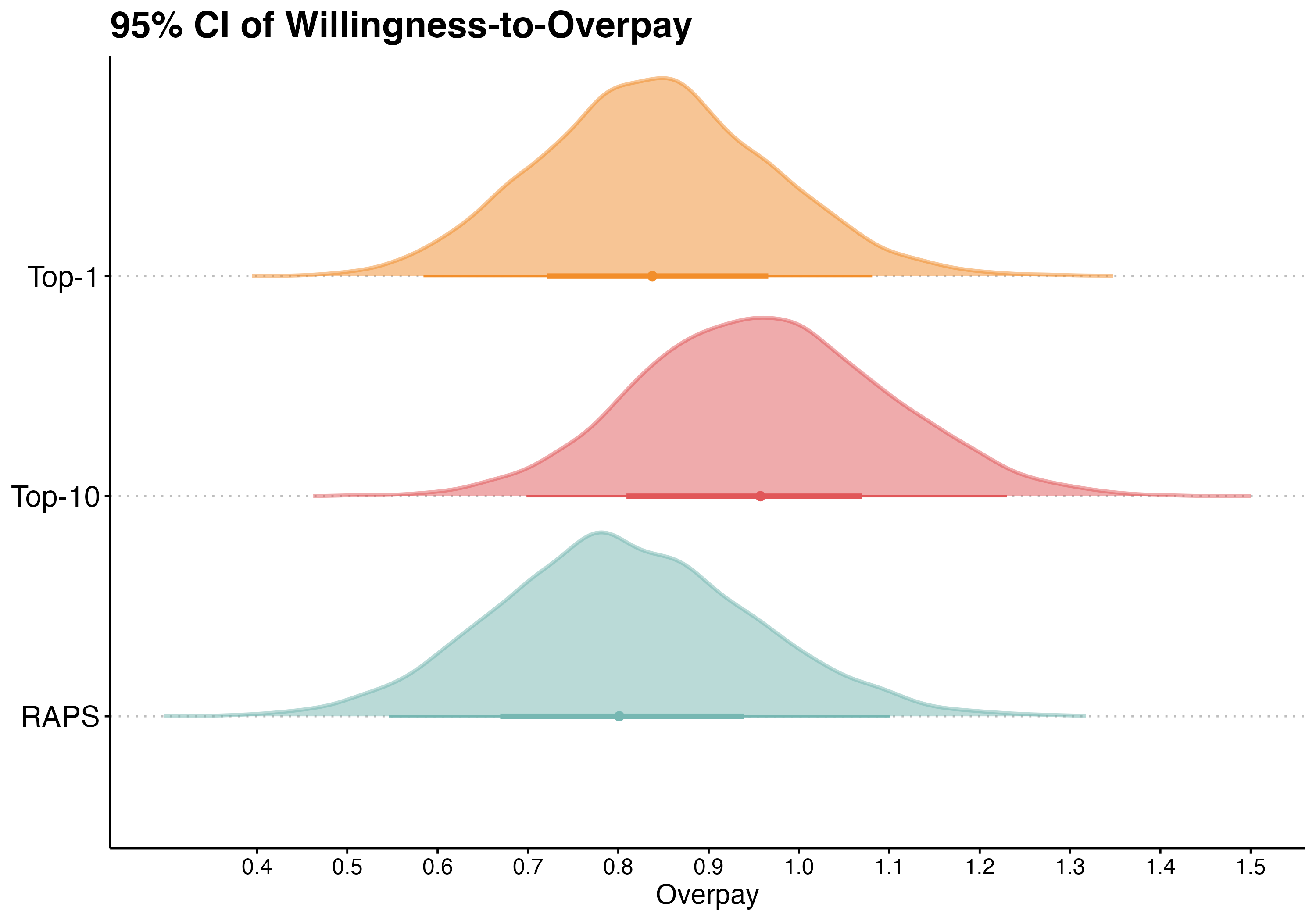}
  \caption[]{
  Bootstrapped 95\% CIs with the median as a point estimate for ``\textit{willingness-to-overpay}'', computed by the difference between the \textit{expected bonus differences} and the willingness-to-pay elicited from participants.
  }
  \label{fig:overpay}
  \Description{
  This figure presents the bootstrapped 95\% confidence intervals for the willingness-to-overpay measure by treatment conditions: Top-1, Top-10, and RAPS.
  }
\end{figure}

We note that this measure is imperfect as a comparator for any individual WTP because we cannot observe the counterfactual performance without a prediction display for each participant individually. In aggregate, however, \autoref{fig:overpay} suggests that participants may over-value the prediction displays to a similar extent across conditions. If anything, despite benefiting more from using the prediction sets, we see weak evidence that \textcolor[HTML]{77B7B3}{\textbf{\texttt{RAPS}}} participants perceive a lower monetary value from the prediction set relative to Top-$k$ predictions. \textcolor[HTML]{77B7B3}{\textbf{\texttt{RAPS}}} participants are willing to overpay by $\$0.8$ on average (CI $[0.54, 1.1]$), slightly less than \textcolor[HTML]{F28E2B}{\textbf{\texttt{Top-1}}} participants ($\$0.84$, CI $[0.598, 1.10]$). Participants find \textcolor[HTML]{E15759}{\textbf{\texttt{Top-10}}} prediction most useful on average ($\$0.96$; CI $[0.7, 1.23]$).

\subsection{Qualitative Analysis of Strategies} \label{sec:strategies}
From our open coding, we find that participants employed different strategies between prediction displays (i.e., ~\autoref{fig:open-codes-results} in Appendix~\ref{sec:appendix-qual-analysis}). In general, most participants reported consulting predictions when decision-making ($\approx$$81\%$ out of 450). Some participants reported first inspecting predictions and assessing their accuracy using their judgment (\textcolor[HTML]{F28E2B}{\textbf{\texttt{Top-1}}}: $\approx$$17\%$ out of 150; \textcolor[HTML]{E15759}{\textbf{\texttt{Top-10}}}: $32\%$ out of 150; \textcolor[HTML]{77B7B3}{\textbf{\texttt{RAPS}}}: $\approx$$39\%$ out of 150), while others report relying first on their own intuitions to form a general idea, and then seeing if it aligned with any predictions (\textcolor[HTML]{F28E2B}{\textbf{\texttt{Top-1}}}: $\approx$$23\%$ out of 150; \textcolor[HTML]{E15759}{\textbf{\texttt{Top-10}}}: $20\%$ out of 150; \textcolor[HTML]{77B7B3}{\textbf{\texttt{RAPS}}}: $\approx$$23\%$ out of 150). Very few participants chose to rely completely on their own intuitions because they find predictions inaccurate or not useful (\textcolor[HTML]{F28E2B}{\textbf{\texttt{Top-1}}}: $\approx$$3\%$ out of 150; \textcolor[HTML]{E15759}{\textbf{\texttt{Top-10}}}: $\approx$$1\%$ out of 150; \textcolor[HTML]{77B7B3}{\textbf{\texttt{RAPS}}}: $\approx$$1\%$ out of 150).
Others reported they would only consult predictions when the image was too hard to recognize and they had no clue (\textcolor[HTML]{F28E2B}{\textbf{\texttt{Top-1}}}: $8\%$ out of 150; \textcolor[HTML]{E15759}{\textbf{\texttt{Top-10}}}: $\approx$$25\%$ out of 150; \textcolor[HTML]{77B7B3}{\textbf{\texttt{RAPS}}}: $10\%$ out of 150). Among participants who reported that they trusted AI prediction(s), some of them in Top-$k$ conditions placed higher trust on predictions with high softmax pseudo-probability (e.g., $\geq50\%$) (\textcolor[HTML]{F28E2B}{\textbf{\texttt{Top-1}}}: $\approx$$30\%$ out of 23; \textcolor[HTML]{E15759}{\textbf{\texttt{Top-10}}}: $\approx$$54\%$ out of 35), while those in \textcolor[HTML]{77B7B3}{\textbf{\texttt{RAPS}}} condition focused more on the top suggestions in the prediction set ($50\%$ out of 26). 

Collectively participants used all features of our interface, reporting that the dropdown ($\approx$$1\%$ out of 599), bottom-up ($\approx$$9\%$ out of 599), and keyword search ($\approx$$13\%$ out of 599) methods were useful in helping them identify a label, and the representative images were informative for aiding understanding of the meaning of labels ($\approx$$14\%$ out of 599). Of participants who viewed prediction displays, those in the \textcolor[HTML]{F28E2B}{\textbf{\texttt{Top-1}}} condition appeared most likely to use the label space hierarchy network to find a better match (\textcolor[HTML]{F28E2B}{\textbf{60\% out of 150}}, compared to \textcolor[HTML]{E15759}{$\boldsymbol{\approx}$\textbf{23\% out of 150}} and \textcolor[HTML]{77B7B3}{\textbf{28\% out of 150}}).

In addition to our open codes, we summarized the proportion of participants who fully relied on predictions (i.e., with all submitted responses selected from predictions). Aligning with the codes described above, we find \textcolor[HTML]{F28E2B}{\textbf{\texttt{Top-1}}} participants relied less on the predictions and used the search tool to find their preferred choice more often for both in-distribution ($\approx$$39\%$ out of 150) and OOD ($\approx$$3\%$ out of 150) images. \textcolor[HTML]{E15759}{\textbf{\texttt{Top-10}}} participants reported relying heavily on predictions for in-distribution images ($\approx$$95\%$ out of 150) and much less for OOD ($25\%$ out of 150). \textcolor[HTML]{77B7B3}{\textbf{\texttt{RAPS}}} participants similarly reported very frequently basing their decisions on a constrained set of predictions in-distribution ($90\%$ out of 150) and slightly less but still quite often for OOD ($48\%$ out of 150).
\section{Discussion}\label{sec:disc}
Conformal prediction has gained recent visibility as a distribution-free approach for quantifying uncertainty in model predictions. Our results suggest that decision-makers may be able to use prediction sets to their advantage in an AI-advised labeling task, but that their advantages over status quo Top-$k$ presentations vary with the properties of the setting. When the model has high accuracy and the test instances are in-distribution, we find that the size of the prediction set is a critical determinant of its utility in our setting. Smaller sets lead to performance on par with or slightly better than Top-$k$ displays while larger ones lead to slightly worse performance. In contrast, when the model encounters unforeseen and challenging OOD instances that can cause Top-$k$ predictions and their associated uncertainty scores to be unreliable (and which are also challenging for the humans), prediction sets offer some comparative advantage OOD, regardless of their set size, at least in a setting like ours where coverage guarantees are maintained.
However, these advantages are relatively small, and may not always outweigh the benefits of instead using the calibration set instances in model training to further improve the classifier.  
We elucidate our main findings and discuss the broader implications of our study.

\paragraph{\textbf{A smaller prediction set derived from a well-calibrated model is generally more useful.}}
Our results suggest that the utility of conformal prediction sets is linked to set size, presumably due to an underlying factor of cognitive load. A large set that embodies more uncertainty requires users to navigate and evaluate many incorrect predictions before finding the correct answer. This increase in cognitive load generally makes prediction sets less effective and may encourage more satisficing~\cite{simon1956rational}, resulting in a lower response accuracy. More rigorous uncertainty quantification does not always lead to better decision-making. Designers of prediction displays should carefully consider the trade-off between set size and adaptiveness when using conformal prediction sets to communicate prediction uncertainty for machine learning models in practical settings.

\paragraph{\textbf{Conformal prediction sets can be more useful for some hard OOD instances.}}
One possible reason prediction sets can exceed in performance for hard OOD stimuli is their adaptiveness. While the covariate shifts used in our experimental setting can severely affect the accuracy of Top-$k$ predictions, adaptive prediction sets under certain shifts can retain coverage by significantly increasing the set size (e.g., \cite{gendler2022adversarially}). While the high coverage of prediction sets for OOD instances does not necessarily equate to the high accuracy of AI-advised humans, as the results for hard instances in~\autoref{fig:acc-results} demonstrate, even large set sizes for OOD instances provided some value to participants. Even if participants did not select the correct label from the set, when given larger prediction sets for hard OOD stimuli, they tended to select labels closer to the correct one, as measured by the shortest path length in the label space hierarchy. It is possible that the additional information about relative uncertainty provided by set size played a role in this advantage. 

\paragraph{\textbf{Withholding predictions of an uncalibrated model may improve decision quality}}
Consistent with prior work on AI-advised decision-making (e.g., \cite{green2019principles, beede2020human, carton2020feature, liu2021understanding}), our results suggest that when a model is well-calibrated and more accurate than humans alone, users with access to its predictions can perform better than without the model, but not as well as the model alone for easier instances. When the model is poorly calibrated, display type affects whether people can perform better by accessing the model predictions. For instance, when the Top-$k$ displays achieved lower coverage of the true label than the conformal guarantee, users sometimes performed worse than they would have done by ignoring the display. We observe qualitative feedback, such as ``I used the Top-$1$ when it was impossible to find what the image was based on my own perception" and ``In cases where I really was unsure and did not know what the picture was, I relied on the AI and went with their label." Decision-making quality, when presented with a constrained set of choices, is correlated with the calibration of prediction information. Only those using RAPS matched the expected performance of having no prediction display in hard OOD instances with larger set sizes. Our results underscore the potential value of treating the detection of instance type as a model's learning problem (e.g., ~\cite{kaur2022codit}).

\subsection{Limitations and Future Work}
It is unclear whether a larger prediction set showing more possible labels can still be effective if the coverage is reduced. A natural follow-up study is to evaluate the efficacy of prediction sets in an OOD setting where coverage is disrupted at varying magnitudes, as well as one where the human retains their relative judgment accuracy and only the AI accuracy is affected. The adaptiveness of the prediction sets~\cite{romano2020, angelopoulos2020uncertainty} enhances a model's resilience to random and common perturbations that can occur in real-world scenarios, such as image corruptions. However, future work may explore the use of intentional adversarial methods that can attack softmax (e.g., FGSM~\cite{goodfellow2014explaining} or PGD~\cite{madry2017towards}). Combined with the finding from~\citet{straitouri2023} that constrained choices can lead to better performance based on smaller sets derived from a calibrated model, we might see an opposite effect to what we observed, where people are less likely to pick a choice from a prediction set that has a lower chance of containing the true label. While our experiment isolated the effect of the conformal prediction set and guarantee without performing other interpretability manipulations to the Top-$k$ conditions, future work might also consider comparing conformal prediction sets to post-hoc calibrated softmax with Top-$k$ displays.

Our experimental design, which does not observe individual participant performance without a prediction display, makes willingness-to-pay an imperfect comparator. Future work should delve into how users perceive the value of prediction displays and conduct within-subject comparisons of both performance and perceived value. Additionally, it is unclear whether the conformal coverage guarantee is prone to misconceptions common to conventional confidence intervals, such as misinterpreting the coverage rate and disregarding labels not included within the set. 
\section{Conclusion}\label{sec:con}
Uncertainty quantification for deep neural networks (NNs) has been challenging due to their black-box nature. We evaluated whether communicating uncertainty via conformal prediction sets can improve human decision-making in AI-advised image labeling. We contribute the results of a large online experiment, in which we evaluate the utility of conformal prediction sets against status quo Top-$k$ predictions across a diverse range of stimuli, varied by in-distribution or OOD, level of difficulty, and set size. Our work shows that when the model is well-calibrated, a smaller prediction set is most beneficial, as larger sets can overwhelm users and decrease performance due to increased cognitive load. However, when the model faces unexpected OOD instances that are also more difficult for humans, larger prediction sets can be more useful than their Top-$k$ counterparts, at least when coverage remains high. Our work sheds light on how conformal prediction sets can be used to rigorously quantify uncertainty for machine learning models in practical applications and outlines potential avenues for future research.
\bibliographystyle{ACM-Reference-Format}
\bibliography{bibliography}

\appendix
\appendix

\section{Further Methodological Details} \label{sec:appendix-methodol}
\begin{figure}[ht]
  \centering
  \includegraphics[width=\linewidth]{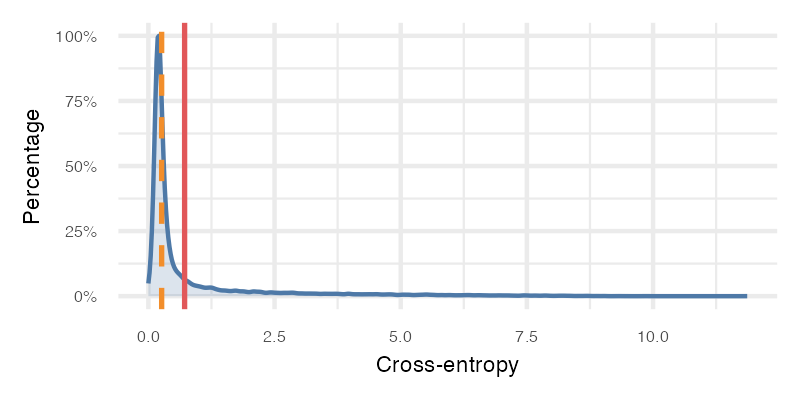}
  \caption[]{
  The distribution of cross-entropy loss observed across in-distribution predictions. The \textcolor[HTML]{F28E2B}{\textbf{\texttt{orange dotted line}}} indicates the median when including images with a set size of one. In contrast, the \textcolor[HTML]{E15759}{\textbf{\texttt{red solid line}}} indicates the median after excluding these images.
  }
  \label{fig:dist-indist-entropy}
  \Description{
    This figure illustrates the distribution of cross-entropy loss for in-distribution predictions. To facilitate comparison of the median cross-entropy values, we use two visual cues: an orange dotted line represents the median when including images with a set size of one, while a red solid line denotes the median after excluding these images. 
    }
\end{figure}

We present the distribution of the cross-entropy loss for stimuli that are in-distribution in \autoref{fig:dist-indist-entropy}. \autoref{tab:all-corruptions-top1-ranking} shows WRN's Top-$k$ prediction accuracy and conformal coverage for in-distribution stimuli and 15 types of covariate shifts. \autoref{tab:task-image-sample-correction} outlines the sampling correction for images with a deterministic set, while \autoref{tab:open-codes-description} explains the open codes used to categorize participants' strategies. Furthermore, \autoref{tab:human-AI-comparison} compares the performance of the Human-AI team, humans working alone, and AI standalone accuracy.

\begin{table}[H]
\caption[]{We evaluated WRN's Top-$\boldsymbol{k}$ prediction accuracy and conformal coverage across 15 types of covariate shifts as discussed by \citet{michaelis2019benchmarking}. For our experimental design, we specifically chose the corruption types marked in bold that reduce WRN's Top-1 prediction accuracy the most as OOD stimuli. The exception is the ``none'' category, which we used as in-distribution stimuli.}
\label{tab:all-corruptions-top1-ranking}
\begin{tabular}{cccc}
\hline
\multicolumn{1}{c}{\multirow{2}{*}{Corruption}} & \multicolumn{3}{c}{Coverage} \\ \cline{2-4} 
\multicolumn{1}{c}{} & \multicolumn{1}{c}{Top-1} & \multicolumn{1}{c}{Top-10} & \multicolumn{1}{c}{RAPS} \\
\hline
\textbf{none} & 0.82 & 0.98 & 0.96 \\
brightness & 0.77 & 0.96 & 0.95 \\
contrast & 0.76 & 0.96 & 0.95 \\
fog & 0.72 & 0.93 & 0.94 \\
jpeg compression & 0.7 & 0.93 & 0.94 \\
pixelate & 0.66 & 0.9 & 0.92 \\
motion blur & 0.57 & 0.83 & 0.88 \\
impulse noise & 0.57 & 0.84 & 0.89 \\
shot noise & 0.56 & 0.83 & 0.89 \\
gaussian noise & 0.55 & 0.83 & 0.89 \\
elastic transform & 0.52 & 0.78 & 0.84 \\
\textbf{defocus blur} & 0.52 & 0.81 & 0.91 \\
\textbf{snow} & 0.51 & 0.79 & 0.86 \\
\textbf{frost} & 0.43 & 0.7 & 0.76 \\
\textbf{zoom blur} & 0.38 & 0.66 & 0.73 \\
\textbf{glass blur} & 0.29 & 0.55 & 0.72 \\
\hline
\end{tabular}
\end{table}

\begin{table}[H]
\centering
\caption{Accuracy achieved by Human-AI teams, humans alone, and AI alone when stimuli were varied by in- or out-of-distribution (i.e., \textit{covariate shift}).}
\label{tab:human-AI-comparison}
\resizebox{\columnwidth}{!}{%
\begin{tabular}{@{}ccccc@{}}
\toprule
\multirow{2}{*}{Condition} & \multirow{2}{*}{Covariate Shift} & \multicolumn{3}{c}{Accuracy} \\ \cmidrule(l){3-5} 
 &  & Human-AI Team & AI Baseline & Human Baseline \\ \midrule
\multirow{2}{*}{Top-1} & in & 0.74 & 0.73 & 0.53 \\
 & out & 0.59 & 0.52 & 0.38 \\ \midrule
\multirow{2}{*}{Top-10} & in & 0.76 & 0.73 & 0.53 \\
 & out & 0.59 & 0.52 & 0.38 \\ \midrule
\multirow{2}{*}{RAPS} & in & 0.73 & 0.73 & 0.53 \\
 & out & 0.63 & 0.52 & 0.38 \\ \bottomrule
\end{tabular}%
}
\end{table}

\begin{table*}
\caption[]{The table demonstrates the sampling correction performed to include images with deterministic set for groups having \texttt{smaller} set size. For each group varied by \textit{Corruption} and \textit{Difficulty}, we compute a \textit{Size Ratio}, which captures the ratio of images with deterministic set (\textit{Size 1 Count}) to uncertain set (\textit{SizeN Count}). \textit{SizeN Keep} shows the number of images with uncertain set whose sizes fall near the median size in that group (i.e., 45\textsuperscript{th} and 55\textsuperscript{th} percentiles). We finally append images with deterministic set (\textit{Size1 Included}) back to the group proportionally according to \textit{Size Ratio}.}
\label{tab:task-image-sample-correction}
\begin{tabular}{cccccccc}
\toprule
\textbf{Corruption} & \textbf{Difficulty} & \textbf{Size1 Count} & \textbf{SizeN Count} & \textbf{Size Ratio} & \textbf{SizeN Keep} & \textbf{Size1 Included} \\
\midrule
none & hard & 539 & 2794 & 0.19 & 749 & \textbf{144} \\
none & easy & 13428 & 2962 & 4.53 & 1778 & \textbf{8060} \\
defocus blur & hard & 331 & 7750 & 0.04 & 870 & \textbf{37} \\
defocus blur & easy & 4571 & 2431 & 1.88 & 385 & \textbf{724} \\
snow & hard & 664 & 7319 & 0.09 & 827 & \textbf{75} \\
snow & easy & 4964 & 2554 & 1.94 & 451 & \textbf{877} \\
frost & hard & 993 & 9073 & 0.11 & 996 & \textbf{109} \\
frost & easy & 2896 & 1590 & 1.82 & 249 & \textbf{454} \\
zoom blur & hard & 722 & 9013 & 0.08 & 1066 & \textbf{85} \\
zoom blur & easy & 2925 & 1914 & 1.53 & 235 & \textbf{359} \\
glass blur & hard & 516 & 9973 & 0.05 & 1108 & \textbf{57} \\
glass blur & easy & 2025 & 1360 & 1.49 & 331 & \textbf{493} \\
\bottomrule
\end{tabular}
\end{table*}

\begin{table*}
\caption[]{Open codes used to summarize participants' strategies.}
\label{tab:open-codes-description}
\begin{tabularx}{\textwidth}{@{}clX@{}}
\toprule
\textbf{Category} & \textbf{Open Codes} & \textbf{Description} \\ 
\midrule
\textbf{1} & UNINFORMATIVE & Content that doesn't provide any significant or relevant information \\
\textbf{2} & HUMAN-AI COMPARISON & Participant analyzes differences and similarities between own's decisions and AI predictions \\
\textbf{3} & HUMAN KNOWLEDGE & Participant makes decisions based solely on own intuition, wisdom, or experience \\
\textbf{4} & HUMAN KNOWLEDGE 1ST \& AI ADVICE 2ND & Participant makes decision based on own's intuition and experience first, supplemented by AI advice or suggestions \\
\textbf{5} & AI ADVICE 1ST \& HUMAN CONFIRM 2ND & Participant seeks advice from AI first and then confirms with own's knowledge \\
\textbf{6} & AI ADVICE WHEN HARD & Participant seeks advice from AI only and specifically in challenging or complex scenarios (e.g., blurry images, etc.) \\
\textbf{7} & TRUST AI PREDICTION(S) & Participant has confidence in the prediction made by AI \\
\textbf{8} & TRUST AI (RANKING OR CONFIDENCE) & Participant relies on AI's ranked suggestions or its confidence level in predictions. Participant mentions looking only at top predictions in RAPS and Top-10, while in Top-1 referring to specific confidence percentages\\
\textbf{9} & TOP PREDICTION FALLACY & Participant expects to see the true answer in top predictions, but they believe the top prediction is wrong \\
\textbf{10} & DISAPPOINTED AT AI & Participant specifically mentions dissatisfaction with AI's performance and does not trust AI's prediction \\
\textbf{11} & SEARCH & Participant performs a general search without specifying \\
\textbf{12} & SEARCH (BY DROPDOWN) & Participant types keywords in the input field and refers to the dropdown for matches \\
\textbf{13} & SEARCH (BY KEYWORD) & Participant clicks ``Search'' to view a hierarchy tree that highlights matching categories \\
\textbf{14} & SEARCH (BY AI LABEL) & Participant clicks an AI-predicted label for a bottom-up search \\
\textbf{15} & SEARCH (HIERARCHY NETWORK) & Participant expands or subtracts nodes/categories and navigates through the provided hierarchy network \\
\textbf{16} & ELIMINATION & Participant tries to narrow down their options (to search in) by elimination \\
\textbf{17} & REPRESENTATIVE IMAGES & Participant compares the task image against the given label-representative images \\
\textbf{18} & COLOR \& SURROUNDINGS & Participant relies on color, shapes, sizes, and surrounding objects to figure out the correct label \\
\bottomrule
\end{tabularx}
\end{table*}

\section{Additional Results}

\subsection{Qualitative Analysis of Strategies} \label{sec:appendix-qual-analysis}
\begin{figure*}
  \centering
  \includegraphics[width=\linewidth]{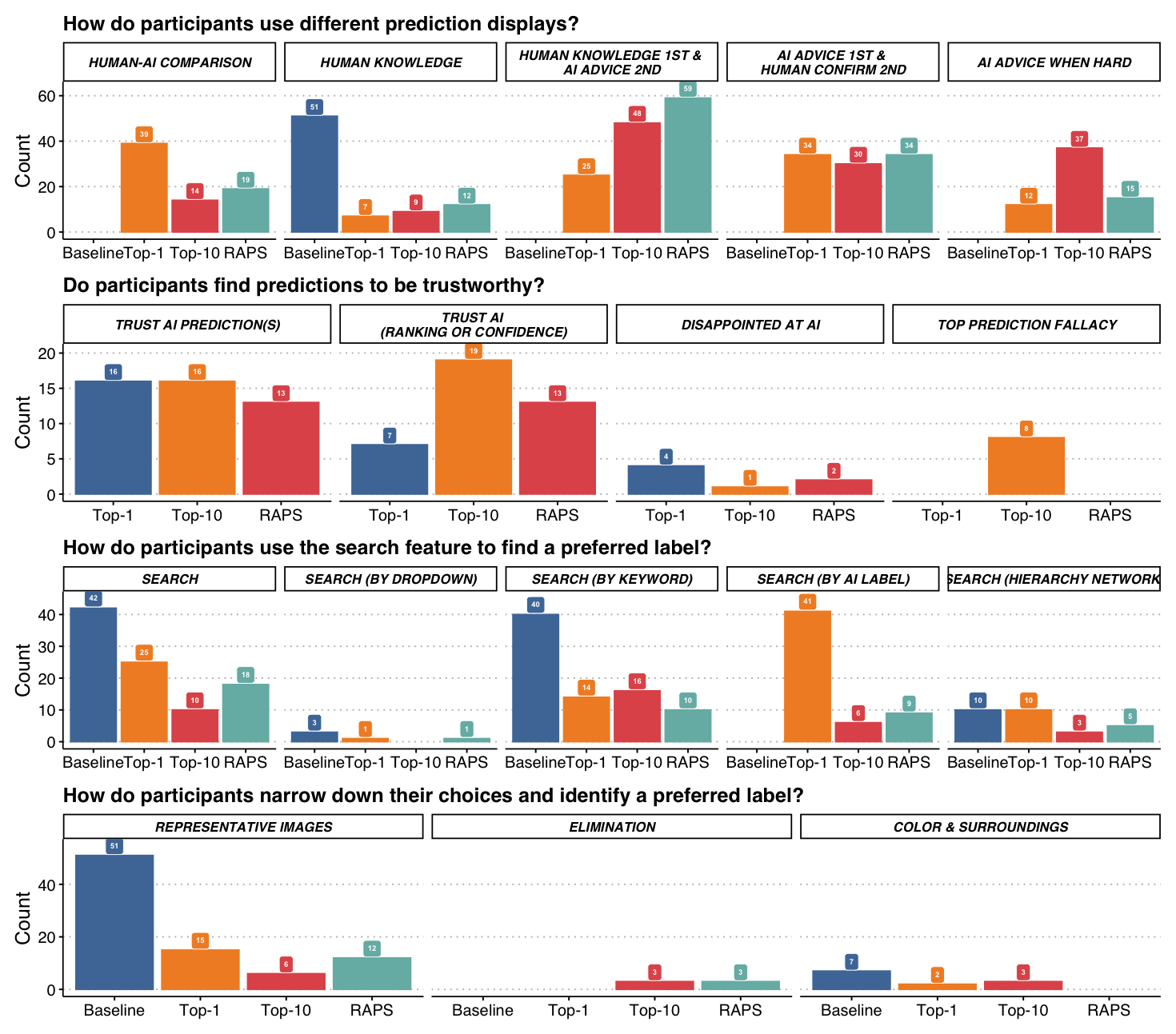}
  \caption[]{Summary of the qualitative results per identified open code.}
  \label{fig:open-codes-results}
  \Description{
  This figure presents a trellis plot summarizing the qualitative results per identified open code. Each row presents a code category with columns showing a barchart summarizing counts of each unique code from participants' strategies by treatment conditions.
  } 
\end{figure*}

\autoref{fig:open-codes-results} presents a trellis plot summarizing the qualitative results per identified open code. Each row presents a code category with columns showing a barchart summarizing counts of each unique code from participants' strategies by treatment conditions.   

\subsection{Shortest Path Length} \label{sec:appendix-sp-results}
The \textit{shortest path (SP) length} quantifies the ``incorrectness'' of participants' labeling choices relative to the correct label in the label space hierarchy network. Similar to accuracy, we analyze the variations of participants' labeling errors by task image types. We report the median of expected predictions for each type of image stimuli with uncertainty quantified as 95\% HDI, holding the trial effect constant at the mean.

\subsubsection{In-distribution}

At a high level, the SP length is, on average, smaller for \texttt{easy} images across conditions. When task images are \texttt{hard}, the SP length is greater for images with \texttt{larger} set sizes across conditions.

When task images are \texttt{easy}, the SP length does not appear to be affected by set size---the distributions of expected SP length closely overlay across conditions (cf. \autoref{fig:sp-results}, A and B). After marginalizing by set sizes, we find that when participants have access to predictions, even when the provided label is wrong, the incorrect label appears to be closer to the correct label in the distance (HPI: $[1.05, 1.6]$) than those of the \textcolor[HTML]{4D79A7}{\textbf{\texttt{baseline}}} condition (HPI: $[2.01, 2.93]$).

When task images are \texttt{hard} but with a \texttt{smaller} average set size of $8$, \autoref{fig:sp-results} (C) shows the SP length is smallest for \textcolor[HTML]{77B7B3}{\textbf{\texttt{RAPS}}} condition with a median of $1.48$ (HPI: $[1.34, 1.64]$). We do not detect differences between the two Top-$k$ conditions, as HPIs closely overlap (\textcolor[HTML]{F28E2B}{\textbf{\texttt{Top-1}}}: $1.8$ HPI: $[1.63, 1.99]$ and \textcolor[HTML]{E15759}{\textbf{\texttt{Top-10}}}: $1.78$ HPI: $[1.57, 2]$). However, when without predictions, the \textcolor[HTML]{4D79A7}{\textbf{\texttt{baseline}}} participants tend to submit more incorrect answers, reflected by the largest SP length of $2.44$ (HPI: $[2.13, 2.84]$).

For \texttt{hard} task images with a \texttt{larger} average set size of $51$, all intervals in \autoref{fig:sp-results} (D) get wider than those in (C), indicating greater variations. Although SP length increased across conditions, we do not detect noticeable differences among conditions with predictions (HPI: $[2.98, 3.42]$). Similarly, the \textcolor[HTML]{4D79A7}{\textbf{\texttt{baseline}}} participants produce more inaccurate responses with a median SP length of $3.83$ (HPI: $[3.31, 4.39]$).

In summary, we find that the SP length mirrors the accuracy in that prediction display that yields higher accuracy, which tends to produce a lower SP length. Between accuracy and SP length, a noticeable difference can be detected from task images that are \texttt{hard} with a \texttt{smaller} set size: although participants in the \textcolor[HTML]{F28E2B}{\textbf{\texttt{Top-1}}} and the \textcolor[HTML]{4D79A7}{\textbf{\texttt{baseline}}} conditions have similar labeling accuracy, the SP length of \textcolor[HTML]{F28E2B}{\textbf{\texttt{Top-1}}} participants is noticeably lower than that of the \textcolor[HTML]{4D79A7}{\textbf{\texttt{baseline}}}. This implies that predictions can help elicit responses closer to the correct label, even if they are wrong. Otherwise, our SP results are mostly consistent with our accuracy results.

\subsubsection{Out-of-distribution}

For \texttt{OOD} task images, the SP length between \texttt{easy} and \texttt{hard} images becomes more contrasting. For \texttt{easy} images, because predictions are still reliable and accurate, we find that the \textcolor[HTML]{F28E2B}{\textbf{\texttt{Top-1}}} condition tends to produce the lowest SP length of $1.35$ (HPI: $[1.24, 1.49]$) than \textcolor[HTML]{E15759}{\textbf{\texttt{Top-10}}} ($1.67$; HPI: $[1.5, 1.9]$) and \textcolor[HTML]{77B7B3}{\textbf{\texttt{RAPS}}} ($1.89$; HPI: $[1.6, 2.32]$), after marginalizing over \texttt{smaller} and \texttt{larger} set sizes (cf. \autoref{fig:sp-results}, E and F).

For \texttt{hard} \texttt{OOD} task images, we see an apparent mirroring effect relative to accuracy. When image stimuli have a \texttt{smaller} average set size of $30$ shown in \autoref{fig:sp-results} (G), \textcolor[HTML]{F28E2B}{\textbf{\texttt{Top-1}}} participants have the lowest accuracy rate and also provide labels that have the highest SP length of $6.22$ (HPI: $[5.77, 6.68]$). \textcolor[HTML]{77B7B3}{\textbf{\texttt{RAPS}}} participants have the highest accuracy for this type of image and produce the lowest SP length of $4.23$ (HPI: $[3.84, 4.65]$). The \textcolor[HTML]{E15759}{\textbf{\texttt{Top-10}}} condition overlaps with \textcolor[HTML]{77B7B3}{\textbf{\texttt{RAPS}}} with an SP length of $4.78$ (HPI: $[4.37, 5.19]$), and both \textcolor[HTML]{E15759}{\textbf{\texttt{Top-10}}} and \textcolor[HTML]{77B7B3}{\textbf{\texttt{RAPS}}} produce lower SP lengths than \textcolor[HTML]{F28E2B}{\textbf{\texttt{Top-1}}} and the \textcolor[HTML]{4D79A7}{\textbf{\texttt{baseline}}} condition ($5.67$; HPI: $[5.22, 6.1]$).

Similarly, for \texttt{hard} \texttt{OOD} task images with a \texttt{larger} average set size of $91$ shown in \autoref{fig:sp-results} (H), \textcolor[HTML]{4D79A7}{\textbf{\texttt{baseline}}} and \textcolor[HTML]{77B7B3}{\textbf{\texttt{RAPS}}} conditions that can help participants achieve the highest labeling accuracy, also support them to give answers closer to the correct label with a median SP length that ranges between $3.03$ and $3.68$, followed by \textcolor[HTML]{F28E2B}{\textbf{\texttt{Top-1}}} ($4.4$; HPI: $[3.99, 4.79]$) and \textcolor[HTML]{E15759}{\textbf{\texttt{Top-10}}} conditions ($4.99$; HPI: $[4.61, 5.41]$).

However, despite that participants of the prediction set have similar accuracy for \texttt{hard} \texttt{OOD} images with \texttt{smaller} and \texttt{larger} sets, we find SP length is smaller for images with \texttt{larger} set sizes. This implies that a \texttt{larger} set size is helpful for \texttt{hard} \texttt{OOD} images, guiding participants to labels closer to the ground truth.

\subsection{Distribution of Willingness-to-Pay} \label{sec:appendix-wtp}
\begin{figure}[ht]
  \centering
  \includegraphics[width=\linewidth]{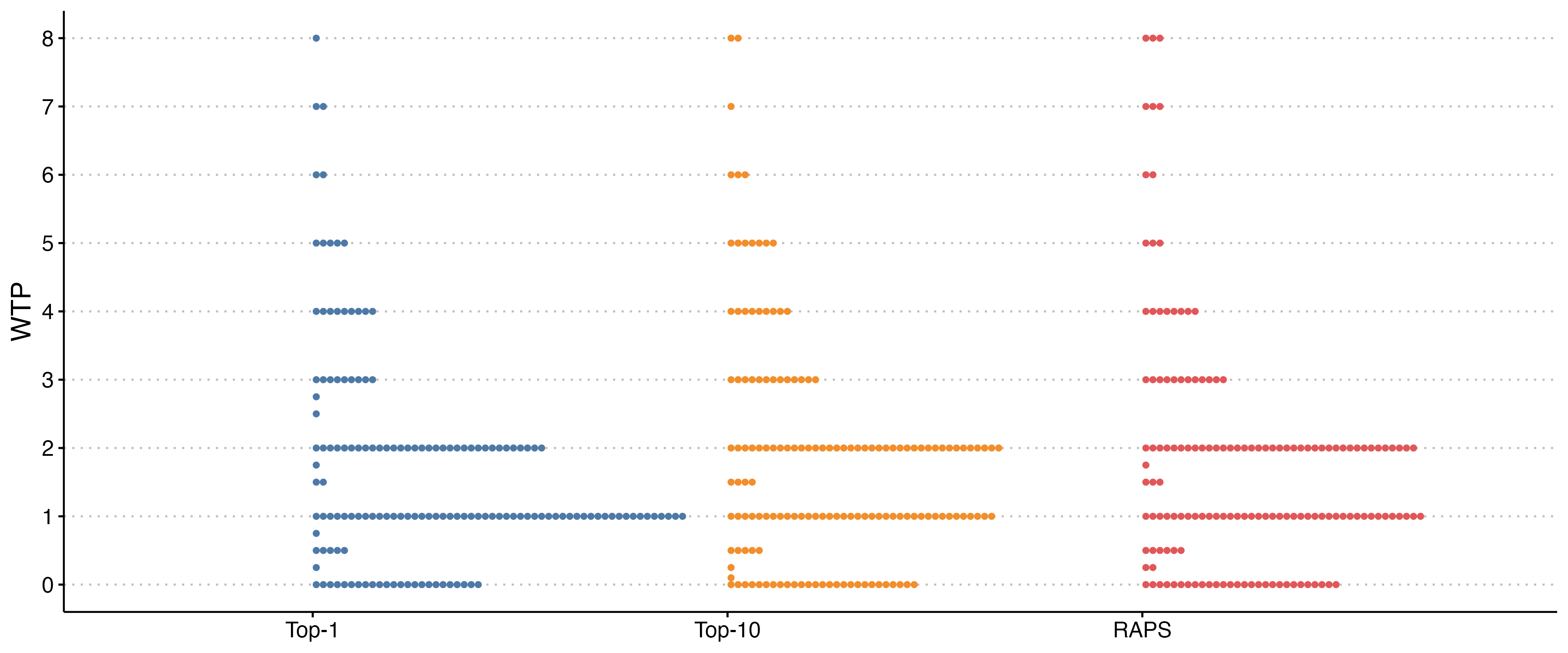}
  \caption[]{Distribution of willingness-to-pay as elicited from participants.}
  \label{fig:wtp-dist}
  \Description{This figure presents the distribution of willingness-to-pay elicited from participants by treatment conditions.}
\end{figure}

\autoref{tab:wtp-descriptive} presents the summary statistics of the elicited willingness-to-pay, while \autoref{fig:wtp-dist} visualizes the distributions using a dot plot.

\begin{table}[ht]
\caption[]{Summary statistics of elicited willingness-to-pay.}
\label{tab:wtp-descriptive}
\begin{tabular}{c c c c c c c c c}
\toprule
\textbf{Condition} & \textbf{Min.} & \textbf{Q1} & \textbf{Q2} & \textbf{Q3} & \textbf{Max.} & \textbf{Mean} & \textbf{SD} \\
\midrule
RAPS & 0 & 1 & 1.0 & 2 & 8 & 1.78 & 1.74 \\
Top-1 & 0 & 1 & 1.0 & 2 & 8 & 1.70 & 1.56 \\
Top-10 & 0 & 1 & 1.5 & 2 & 8 & 1.84 & 1.67 \\
\bottomrule
\end{tabular}
\end{table}

\end{document}